\documentclass[english,aps,manuscript]{revtex4}
\usepackage[T1]{fontenc}
\usepackage[latin9]{inputenc}
\usepackage[a4paper]{geometry}
\usepackage{amsmath}
\usepackage{amssymb}
\usepackage{graphicx}
\usepackage{esint}

\makeatletter
\@ifundefined{textcolor}{}
{%
 \definecolor{BLACK}{gray}{0}
 \definecolor{WHITE}{gray}{1}
 \definecolor{RED}{rgb}{1,0,0}
 \definecolor{GREEN}{rgb}{0,1,0}
 \definecolor{BLUE}{rgb}{0,0,1}
 \definecolor{CYAN}{cmyk}{1,0,0,0}
 \definecolor{MAGENTA}{cmyk}{0,1,0,0}
 \definecolor{YELLOW}{cmyk}{0,0,1,0}
}

\@ifundefined{showcaptionsetup}{}{%
 \PassOptionsToPackage{caption=false}{subfig}}
\usepackage{subfig}
\makeatother

\usepackage{babel}
\begin{document}

\title{Coupling vortex dynamics with collective excitations in Bose-Einstein
Condensates}

\author{R. P. Teles, V. S. Bagnato and F. E. A. dos Santos}

\affiliation{Instituto de Física de São Carlos, USP, Caixa Postal 369, 13560-970
São Carlos, São Paulo, Brazil}
\begin{abstract}
Here we analyze the collective excitations as well as the expansion
of a trapped Bose-Einstein condensate with a vortex line at its center.
To this end, we propose a variational method where the variational
parameters have to be carefully chosen in order to produce reliable
results. Our variational calculations agree with numerical simulations
of the Gross-Pitaevskii equation. The system considered here turns
out to exhibit four collective modes of which only three can be observed
at a time depending of the trap anisotropy. We also demonstrate that
these collective modes can be excited using well stablished experimental
methods such as modulation of the s-wave scattering length.
\end{abstract}
\maketitle

\section{Introduction}

{\indent}

In this work, we are interested in the dynamics of a trapped Bose-Einstein
condensate (BEC) containing a line vortex at its center. Here we are
particularly interested in obtaining the collective oscillation modes
of the system which couples the vortex core oscillations with the
oscillations of the condensate external dimensions. The interest in
this problem is motivated by the fact that these oscillations can
be measured in the laboratory by moving the atomic cloud out of its
equilibrium configuration by using the Feshbach resonance in order
to modulate the scattering length \cite{cm1,c-ex1,perez2,rev1,rev3}.
These oscillations are also studied in other physical systems such
as two-species condensates \cite{2-comp}, BCS-BEC crossover \cite{cm2,cm3,cm5},
and superfluid Helium \cite{cm4}. From the theoretical point of view,
we are interested on how the size of the vortex core oscillates with
respect to the external dimensions of the cloud. The mode with the
smallest oscillation frequency is the quadrupole mode which occurs
when the longitudinal and radial sizes of condensate oscillate out
phase. The breathing mode requires more energy to be excited since
the change in the density of the atomic cloud imposes a greater resistance
against deviation from its equilibrium configuration than in the case
of quadrupole excitations \cite{pethick,pit-str}.

In Refs. \cite{fetter1,vi2,sa-nomalmodes}, the dynamics of normal
modes for a single vortex has been studied using hydrodynamic models,
which focus on the vortex motion with respect to the center-of-mass
of the condensate. This concept was also used in the case of multicomponent
Bose-Einstein condensates \cite{2-compvor}.

\begin{figure}
\centering

\includegraphics{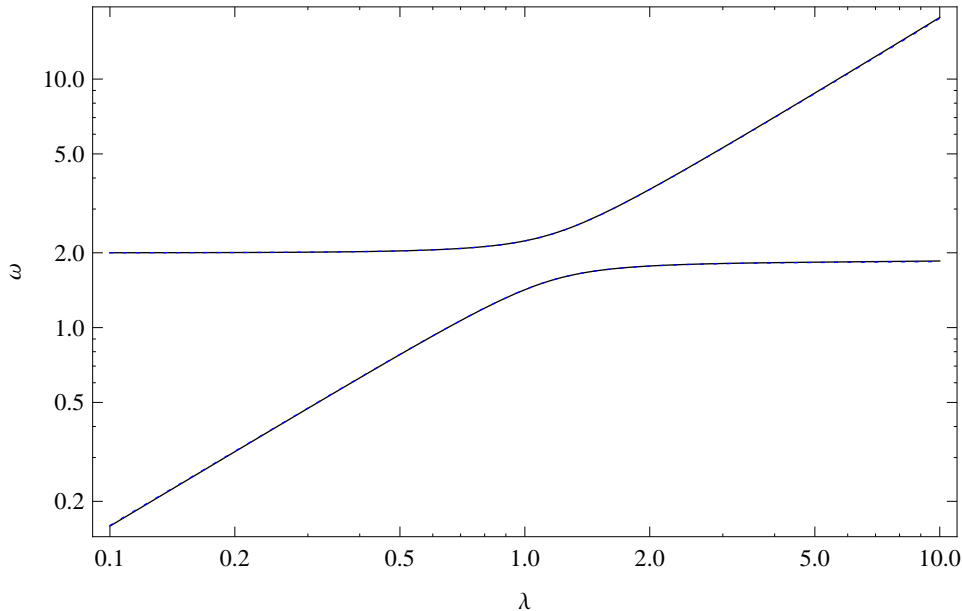}

\caption{(Color online) Oscillation frequencies from Gaussian Ansatz without
taking into account the independent variation of the vortex core size.
Upper lines correspond the frequencies of the breathing mode as a
function of the harmonic trap anisotropy, whereas lower lines represent
the frequencies of the quadrupole mode. Solid (black) lines correspond
to a vortex-free Gaussian-profile while dotted (blue) lines describe
a profile with a singly charged vortex. Note that $\varpi$ is normalized
by the frequency of the radial direction $\omega_{\rho}$.}

\label{fig1}
\end{figure}

Preliminary calculations using a variational calculation with a Gaussian
Ansatz, which does not take into account the independent variation
of the vortex core size \cite{perez1,perez2,2-comp,2-compvor}, shows
a small shift in the frequencies of the aforementioned modes (Figure
\ref{fig1}). This shift has already been obtained via a hydrodynamic
approximation in Refs. \cite{pit-str} and \cite{normalmodes}.
 Thus we can expect the frequency of the monopole (breathing) mode
to decrease while the quadrupole frequency increases in the presence
of the vortex.

To calculate the dynamics of a vortex with charge $\ell$ in a more
consistent way with the physical reality, which allows for the coupling
between vortex core and the external dimensions of condensate, we
could naïvely use a Thomas-Fermi (TF) Ansatz \cite{vi1}
\begin{equation}
\psi\left(\rho,\varphi,z,t\right)=A\left(t\right)\left[\frac{\rho^{2}}{\rho^{2}+\xi\left(t\right)^{2}}\right]^{\frac{1}{2}}\sqrt{1-\frac{\rho^{2}}{R_{\rho}\left(t\right)^{2}}-\frac{z^{2}}{R_{z}\left(t\right)^{2}}}\exp\left[i\ell\varphi+iB_{\rho}\left(t\right)\frac{\rho^{2}}{2}+iB_{z}\left(t\right)\frac{z^{2}}{2}\right],\label{eq:TF-A}
\end{equation}
and then calculate the equations of motion for the five variational
parameters ($\xi$, $R_{\rho}$, $R_{z}$, $B_{\rho}$, $B_{z}$).
Following these calculations, the equations of motion would be linearized.
For the Ansatz (\ref{eq:TF-A}), this procedure leads to imaginary
frequencies which are not consistent with the stable configuration
where a singly charged ($\ell=1$) vortex resides at the center of
the condensate. The linearized equations of motion can be written
in a matrix form according to
\begin{equation}
M\ddot{\delta}+V\delta=0,\label{eq:2}
\end{equation}
where $\delta$ is the vector with components given by deviations
of the variational parameters from their equilibrium values. The solution
of (\ref{eq:2}) is a linear combination of oscillatory modes whose
oscillation frequencies obey the equation 
\begin{equation}
\prod_{n}\varpi_{n}^{2}=\det\left(M^{-1}V\right)=\frac{\det V}{\det M}.\label{eq:w2}
\end{equation}

In order to ensure that all frequencies $\varpi_{n}$ are real, we
must have $\det V/\det M>0$. We know that $\det V>0$ since its sign
reflects the sign of the variational parameters which represents the
external dimensions of the cloud in the stationary situation. Therefore
$\det M$ must also be positive. In the case of Ansatz (\ref{eq:TF-A})
with $\ell=1$, such conditions are not satisfied since $\det M<0$,
which indicates that there is something wrong with Ansatz (\ref{eq:TF-A}).
In previous works \cite{2-compvor,michele,sa-nomalmodes,rpteles1},
since the authors did not consider the size of the vortex core as
a variational parameter, this problem did not appear. Indeed, the
problem relies on the fact that the phase of the wave function have
to be modified.

In Section II, the necessary requirements for the wave function phase
are discussed in order to give support to our variational method.
Section III has the calculation based on the new Ansatz and the corresponding
equations of motion are obtained. The collective modes considering
the coupling between vortex and atomic cloud are obtained via linearization
of the equations of motion, thus resulting in new collective oscillations
(section IV). In section V, we showed that such excitation modes can
be excited using the scattering length modulation. The free expansion
was also calculated in order to complement a previous work \cite{rpteles1}.
Finally, section VII contains the conclusions on our subject of study.

\section{Wave-function phase}

{\indent}

We start with the Lagrangian density,
\begin{equation}
\mathcal{L}=\frac{i\hbar}{2}\left(\psi^{*}\frac{\partial\psi}{\partial t}-\psi\frac{\partial\psi^{*}}{\partial t}\right)-\frac{\hbar^{2}}{2m}\left|\nabla\psi\right|^{2}-V\left(\mathbf{r}\right)\left|\psi\right|^{2}-\frac{g}{2}\left|\psi\right|^{4},\label{eq:2-1}
\end{equation}

whose extremization leads to the Gross-Pitaevskii equation (GPE):

\begin{equation}
i\hbar\frac{\partial\psi}{\partial t}=\left[-\frac{\hbar^{2}}{2m}\nabla^{2}+V\left(\mathbf{r}\right)+g\left|\psi\right|^{2}\right]\psi,
\end{equation}
where $V\left(\mathbf{r}\right)=\frac{1}{2}m\omega_{\rho}^{2}\left(\rho^{2}+\lambda^{2}z^{2}\right)$
is an external potential, the trap anisotropy is $\lambda=\omega_{z}/\omega_{\rho}$,
and $g$ is the coupling constant. The complex field $\psi\left(\mathbf{r},t\right)$
can be written as an amplitude profile multiplied by a respective
phase, as follows: 

\begin{equation}
\psi\left(\mathbf{r},t\right)=f\left(w_{l},\mathbf{r}\right)e^{iS\left(\chi_{l},\mathbf{r}\right)},\label{eq:2-2}
\end{equation}
where

\begin{equation}
S\left(\chi_{l},\mathbf{r}\right)=\ell\varphi+\sum_{l}\chi_{l}\phi_{l}(\mathbf{r}).\label{eq:2-3}
\end{equation}
We denoted both, $w_{l}=w_{l}\left(t\right)$ and $\chi_{l}=\chi_{l}\left(t\right)$,
respectively, as the amplitude and phase variational parameters. In
principle, $\left\{ \phi_{l}\left(\mathbf{r}\right)\right\} $ should
be a complete set of functions but in our present approximation, we
use only a representative incomplete set of functions. Substituting
(\ref{eq:2-2}) and (\ref{eq:2-3}) into (\ref{eq:2-1}), the Lagrangian
$L=\int\mathcal{L}d^{3}\mathbf{r}$ becomes

\begin{equation}
L=-\hbar\sum_{l}\dot{\chi_{l}}\int d^{3}\mathbf{r}f^{2}\phi_{l}-\frac{\hbar^{2}}{2m}\sum_{l}\chi_{l}^{2}\int d^{3}\mathbf{r}f^{2}\left|\nabla\phi_{l}\right|^{2}-\int d^{3}\mathbf{r}\left(\frac{\hbar^{2}}{2m}\left|\nabla f\right|^{2}+Vf^{2}+\frac{g}{2}f^{4}\right).
\end{equation}

In order to account for the dynamics of all three variational parameter
in $f$ we include a variational phase which also contains three variational
parameters. This way we chose the following trial function: 
\begin{equation}
S(\rho,z,t)=\ell\varphi+B_{\rho}(t)\frac{\rho^{2}}{2}+C(t)\frac{\rho^{4}}{4}+B_{z}(t)\frac{z^{2}}{2}.\label{eq:fase 2}
\end{equation}
As the superfluid current is connected to the density variation, it
is desirable that both, amplitude and phase, have the same number
of variational parameters. The Ansatz (\ref{eq:fase 2}) also leads
to linearized equations of motion (\ref{eq:2}) with $\det M>0$ which
is consistent with the stability of the condensate with a singly charged
vortex at its center.

\section{Equations of motion}

{\indent}

Now we correct the Thomas-Fermi Ansatz according to the discussion
in section II. This leads to the following trial function:
\begin{align}
\psi\left(\mathbf{r},t\right)= & \sqrt{\frac{N}{R_{\rho}(t)^{2}R_{z}(t)A_{0}\left[\xi\left(t\right)/R_{\rho}\left(t\right)\right]}}\left[\frac{\rho^{2}}{\rho^{2}+\xi\left(t\right)^{2}}\right]^{\frac{\ell}{2}}\sqrt{1-\frac{\rho^{2}}{R_{\rho}\left(t\right)^{2}}-\frac{z^{2}}{R_{z}\left(t\right)^{2}}}\nonumber \\
 & \times\exp\left[i\ell\varphi+iB_{\rho}\left(t\right)\frac{\rho^{2}}{2}+iC(t)\frac{\rho^{4}}{4}+iB_{z}\left(t\right)\frac{z^{2}}{2}\right],\label{eq:NTF}
\end{align}
with
\begin{align}
A_{0}(\alpha)= & \frac{2\pi^{3/2}\left(\ell\right)!}{15\alpha^{2\ell}\left(\frac{3}{2}+\ell\right)!}\left[\left(3+2\ell\alpha^{2}\right){}_{2}F_{1}\left(\ell,1+\ell;\frac{5}{2}+\ell;-\frac{1}{\alpha^{2}}\right)\right.\nonumber \\
 & \left.-2\ell\left(1+\alpha^{2}\right){}_{2}F_{1}\left(1+\ell,1+\ell;\frac{5}{2}+\ell;-\frac{1}{\alpha^{2}}\right)\right],
\end{align}
where, for simplicity we define $\alpha\left(t\right)=\xi\left(t\right)/R_{\rho}\left(t\right)$,
$_{p}F_{q}\left(a_{1},\ldots,a_{p};b_{1},\ldots,b_{q};x\right)$ are
the hypergeometric functions, $\xi(t)$ is the size of the vortex
core, $R_{\rho}(t)$ is the condensate size in radial direction ($\hat{\rho}$),
and $R_{z}(t)$ is the condensate size in axial direction ($\hat{z}$).
The wave function (\ref{eq:NTF}) has integration domain defined by
$1-\frac{\rho^{2}}{R_{\rho}^{2}}-\frac{z^{2}}{R_{z}^{2}}\geq0$, where
the wave function is approximately an inverted parabola (TF-shape),
except for the central vortex. The trapping potential shape sets the
condensate dimensions. To organize our calculations, we split the
Lagrangian so that it is a sum $L=L_{time}+L_{kin}+L_{pot}+L_{int}$
of the following terms:
\begin{align}
L_{time}= & \frac{i\hbar}{2}\int d^{3}\mathbf{r}\left[\psi^{*}\left(\mathbf{r},t\right)\frac{\partial\psi\left(\mathbf{r},t\right)}{\partial t}-\psi\left(\mathbf{r},t\right)\frac{\partial\psi^{*}\left(\mathbf{r},t\right)}{\partial t}\right]\nonumber \\
= & -\frac{N\hbar}{2}\left(D_{1}\dot{B_{\rho}}R_{\rho}^{2}+D_{2}\dot{B_{z}}R_{z}^{2}+\frac{1}{2}D_{3}\dot{C}R_{\rho}^{4}\right),\\
L_{kin}= & -\frac{\hbar^{2}}{2m}\int d^{3}\mathbf{r}\left[\nabla\psi^{*}\left(\mathbf{r},t\right)\right]\left[\nabla\psi\left(\mathbf{r},t\right)\right]\nonumber \\
= & -\frac{N\hbar^{2}}{2m}\left[D_{1}B_{\rho}^{2}R_{\rho}^{2}+D_{2}B_{z}^{2}R_{z}^{2}+2D_{3}B_{\rho}CR_{\rho}^{4}+R_{\rho}^{-2}\left(\ell^{2}D_{4}+D_{5}\right)+D_{6}C^{2}R_{\rho}^{6}\right],\\
L_{pot}= & -\frac{1}{2}m\omega_{\rho}^{2}\int d^{3}\mathbf{r}\left(\rho^{2}+\lambda^{2}z^{2}\right)\psi^{*}\left(\mathbf{r},t\right)\psi\left(\mathbf{r},t\right)\nonumber \\
= & -\frac{N}{2}m\omega_{\rho}^{2}\left(D_{1}R_{\rho}^{2}+\lambda^{2}D_{2}R_{z}^{2}\right),\\
L_{int}= & -\frac{g}{2}\int d^{3}\mathbf{r}\left[\psi^{*}\left(\mathbf{r},t\right)\psi\left(\mathbf{r},t\right)\right]^{2}\nonumber \\
= & -\frac{N^{2}gD_{7}}{2R_{\rho}^{2}R_{z}},
\end{align}
with the functions $D_{i}\left(\alpha\right)$ given by
\begin{align}
D_{1}\left(\alpha\right)= & A_{0}\left(\alpha\right)^{-1}\frac{2\pi^{3/2}\left(1+\ell\right)!}{21\alpha^{2\ell}\left(\frac{5}{2}+\ell\right)!}\left[\left(3+2\ell\alpha^{2}\right){}_{2}F_{1}\left(\ell,2+\ell;\frac{7}{2}+\ell;-\frac{1}{\alpha^{2}}\right)\right.\nonumber \\
 & \left.-2\ell\left(1+\alpha^{2}\right){}_{2}F_{1}\left(1+\ell,2+\ell;\frac{7}{2}+\ell;-\frac{1}{\alpha^{2}}\right)\right],\\
D_{2}\left(\alpha\right)= & A_{0}\left(\alpha\right)^{-1}\frac{\pi^{3/2}\left(\ell\right)!}{4\alpha^{2\ell}\left(\frac{7}{2}+\ell\right)!}\left[\left(7+2\ell\right){}_{2}F_{1}\left(\ell,1+\ell;\frac{7}{2}+\ell;-\frac{1}{\alpha^{2}}\right)\right.\nonumber \\
 & \left.-\left(5+2\ell\right){}_{3}F_{2}\left(\ell,1+\ell,\frac{7}{2}+\ell;\frac{5}{2}+\ell,\frac{9}{2}+\ell;-\frac{1}{\alpha^{2}}\right)\right],\\
D_{3}\left(\alpha\right)= & A_{0}\left(\alpha\right)^{-1}\frac{2\pi^{3/2}\left(2+\ell\right)!}{27\alpha^{2\ell}\left(\frac{7}{2}+\ell\right)!}\left[\left(3+2\ell\right){}_{2}F_{1}\left(\ell,3+\ell;\frac{9}{2}+\ell;-\frac{1}{\alpha^{2}}\right)\right.\nonumber \\
 & \left.-2\ell\left(1+\alpha^{2}\right){}_{2}F_{1}\left(1+\ell,3+\ell;\frac{9}{2}+\ell;-\frac{1}{\alpha^{2}}\right)\right],\\
D_{4}\left(\alpha\right)= & A_{0}\left(\alpha\right)^{-1}\frac{2\pi^{3/2}\left(\ell-1\right)!}{3\alpha^{2\ell}\left(\frac{1}{2}+\ell\right)!}\left[\left(1-2\ell\alpha^{2}\right){}_{2}F_{1}\left(\ell,2+\ell;\frac{3}{2}+\ell;-\frac{1}{\alpha^{2}}\right)\right.\nonumber \\
 & \left.+2\ell\left(1+\alpha^{2}\right){}_{2}F_{1}\left(1+\ell,2+\ell;\frac{3}{2}+\ell;-\frac{1}{\alpha^{2}}\right)\right],\\
D_{5}\left(\alpha\right)= & A_{0}\left(\alpha\right)^{-1}\frac{2\pi^{3/2}\left(\ell-1\right)!}{9\alpha^{2\ell}\left(\frac{1}{2}+\ell\right)!}\left[\left(3+2\ell\alpha^{2}\right){}_{2}F_{1}\left(\ell,\ell;\frac{3}{2}+\ell;-\alpha^{2}\right)\right.\nonumber \\
 & \left.-2\ell\left(1+\alpha^{2}\right){}_{2}F_{1}\left(\ell,1+\ell;\frac{3}{2}+\ell;-\frac{1}{\alpha^{2}}\right)\right],\\
D_{6}\left(\alpha\right)= & A_{0}\left(\alpha\right)^{-1}\frac{2\pi^{3/2}\left(3+\ell\right)!}{33\alpha^{2\ell}\left(\frac{9}{2}+\ell\right)!}\left[\left(3+2\ell\alpha^{2}\right){}_{2}F_{1}\left(\ell,4+\ell;\frac{11}{2}+\ell;-\frac{1}{\alpha^{2}}\right)\right.\nonumber \\
 & \left.-2\ell\left(1+\alpha^{2}\right){}_{2}F_{1}\left(1+\ell,4+\ell;\frac{11}{2}+\ell;-\frac{1}{\alpha^{2}}\right)\right],\\
D_{7}\left(\alpha\right)= & A_{0}\left(\alpha\right)^{-2}\frac{2\pi^{3/2}\left(2\ell\right)!}{\alpha^{4\ell}\left(\frac{7}{2}+\ell\right)!}{}_{2}F_{1}\left(2\ell,1+2\ell;\frac{9}{2}+2\ell;-\frac{1}{\alpha^{2}}\right).
\end{align}
For simplicity we can scale the variational parameters of the Lagrangian
as well as the time in order to make them dimensionless,
\begin{align*}
R_{\rho}(t)\rightarrow & a_{osc}r_{\rho}(t),\\
R_{z}(t)\rightarrow & a_{osc}r_{z}(t),\\
\xi(t)\rightarrow & a_{osc}r_{\xi}(t),\\
B_{\rho}(t)\rightarrow & a_{osc}^{-2}\beta_{\rho}(t),\\
B_{z}(t)\rightarrow & a_{osc}^{-2}\beta_{z}(t),\\
C(t)\rightarrow & a_{osc}^{-4}\zeta(t),\\
t\rightarrow & \omega_{\rho}^{-1}\tau,
\end{align*}
where the harmonic oscillator length is $a_{osc}=\sqrt{\hbar/m\omega_{\rho}}$
and the dimensionless interaction parameter is $\gamma=Na_{s}/a_{osc}$.
Thus the Lagrangian becomes
\begin{align}
L= & -\frac{N\hbar\omega_{\rho}}{2}\left[D_{1}r_{\rho}^{2}\left(\dot{\beta_{\rho}}+\beta_{\rho}^{2}+1\right)+D_{2}r_{z}^{2}\left(\dot{\beta_{z}}+\beta_{z}^{2}+\lambda^{2}\right)\right.\nonumber \\
 & \left.+D_{3}r_{\rho}^{4}\left(\frac{1}{2}\dot{\zeta}+2\beta_{\rho}\zeta\right)+\ell^{2}r_{\rho}^{-2}\left(D_{4}+D_{5}\right)+D_{6}\zeta^{2}r_{\rho}^{6}+D_{7}\frac{4\pi\gamma}{r_{\rho}^{2}r_{z}}\right].\label{eq:Lagrangian}
\end{align}
The Euler-Lagrange equations
\begin{align}
\frac{d}{dt}\left(\frac{\partial L}{\partial\dot{q_{i}}}\right)-\frac{\partial L}{\partial q_{i}}= & 0,
\end{align}
for each one of the six variational parameters from Lagrangian (\ref{eq:Lagrangian})
lead to the six differential equations:
\begin{align}
\beta_{\rho}-\frac{\dot{r_{\rho}}}{r_{\rho}}-\frac{D_{1}^{\prime}\dot{\alpha}}{2D_{1}}+\frac{D_{3}r_{\rho}^{2}\zeta}{D_{1}} & =0,\label{eq:1a}\\
\beta_{z}-\frac{\dot{r_{z}}}{r_{z}}-\frac{D_{2}^{\prime}\dot{\alpha}}{2D_{2}} & =0,\label{eq:1b}\\
\zeta-\frac{D_{3}\dot{r_{\rho}}}{D_{6}r_{\rho}}-\frac{D_{3}\dot{\alpha}}{4D_{6}r_{\rho}^{2}}+\frac{D_{3}\beta_{\rho}}{D_{6}r_{\rho}^{2}} & =0,\label{eq:1c}\\
D_{1}r_{\rho}\left(\dot{\beta_{\rho}}+\beta_{\rho}^{2}+1\right)+D_{3}r_{\rho}^{3}\left(\dot{\zeta}+4\beta_{\rho}\zeta\right)-\nonumber \\
\frac{\ell^{2}}{r_{\rho}^{3}}\left(D_{4}+D_{5}\right)+3D_{6}\zeta^{2}r_{\rho}^{5}-D_{7}\frac{4\pi\gamma}{r_{\rho}^{3}r_{z}} & =0,\label{eq:1d}\\
D_{2}r_{z}\left(\dot{\beta_{z}}+\beta_{z}^{2}+\lambda^{2}\right)-D_{7}\frac{2\pi\gamma}{r_{\rho}^{2}r_{z}^{2}} & =0,\label{eq:1e}\\
D_{1}^{\prime}r_{\rho}^{2}\left(\dot{\beta_{\rho}}+\beta_{\rho}^{2}+1\right)+D_{2}^{\prime}r_{z}^{2}\left(\dot{\beta_{z}}+\beta_{z}^{2}+\lambda^{2}\right)+D_{3}^{\prime}r_{\rho}^{4}\left(\frac{1}{2}\dot{\zeta}+2\beta_{\rho}\zeta\right)+\nonumber \\
\frac{\ell^{2}}{r_{\rho}^{2}}\left(D_{4}^{\prime}+D_{5}^{\prime}\right)+D_{6}^{\prime}\zeta^{2}r_{\rho}^{6}-D_{7}^{\prime}\frac{4\pi\gamma}{r_{\rho}^{2}r_{z}} & =0.\label{eq:1f}
\end{align}
 {\indent}

Solving these equations for the parameters in the wave function phase,
we have:
\begin{align}
\beta_{\rho}= & \frac{\dot{r_{\rho}}}{r_{\rho}}+F_{1}\dot{\alpha},\label{eq:2a}\\
\beta_{z}= & \frac{\dot{r_{z}}}{r_{z}}+F_{2}\dot{\alpha},\label{eq:2b}\\
\zeta= & F_{3}\frac{\dot{\alpha}}{r_{\rho}^{2}},\label{eq:2c}
\end{align}
where
\begin{align}
F_{1}= & \frac{D_{3}^{\prime}D_{3}-2D_{1}^{\prime}D_{6}}{4\left(D_{3}^{2}-D_{1}D_{6}\right)},\\
F_{2}= & \frac{D_{2}^{\prime}}{2D_{2}},\\
F_{3}= & \frac{2D_{1}^{\prime}D_{3}-D_{1}D_{3}^{\prime}}{4\left(D_{3}^{2}-D_{1}D_{6}\right)}.
\end{align}
Replacing (\ref{eq:2a}), (\ref{eq:2b}), and (\ref{eq:2c}) into
equations (\ref{eq:1d}), (\ref{eq:1e}), and (\ref{eq:1f}), we reduce
our six coupled equations to only three, which are given by:
\begin{align}
D_{1}\left(\ddot{r_{\rho}}+r_{\rho}\right)+G_{1}r_{\rho}\ddot{\alpha}+G_{2}r_{\rho}\dot{\alpha}^{2}+G_{3}\dot{r_{\rho}}\dot{\alpha}-G_{4}\frac{\ell^{2}}{r_{\rho}^{3}}-D_{7}\frac{4\pi\gamma}{r_{\rho}^{3}r_{z}} & =0,\label{eq:2d}\\
D_{2}\left(\ddot{r_{z}}+\lambda^{2}r_{z}\right)+G_{5}r_{z}\ddot{\alpha}+G_{6}r_{z}\dot{\alpha}^{2}+G_{7}\dot{r_{z}}\dot{\alpha}-D_{7}\frac{2\pi\gamma}{r_{\rho}^{2}r_{z}^{2}} & =0,\label{eq:2e}\\
D_{1}^{\prime}r_{\rho}\left(\ddot{r_{\rho}}+r_{\rho}\right)+D_{2}^{\prime}r_{z}\left(\ddot{r_{z}}+\lambda^{2}r_{z}\right)+\left(G_{8}r_{\rho}^{2}+G_{9}r_{z}^{2}\right)\ddot{\alpha}+\left(G_{10}r_{\rho}^{2}+G_{11}r_{z}^{2}\right)\dot{\alpha}^{2}\nonumber \\
+\left(G_{12}r_{\rho}\dot{r_{\rho}}+G_{13}r_{z}\dot{r_{z}}\right)\dot{\alpha}+G_{14}\frac{\ell^{2}}{r_{\rho}^{2}}+D_{7}^{\prime}\frac{4\pi\gamma}{r_{\rho}^{2}r_{z}} & =0,\label{eq:2f}
\end{align}
with
\begin{align}
G_{1}= & D_{1}F_{1}+D_{3}F_{3},\\
G_{2}= & D_{1}\left(F_{1}^{2}+F_{1}^{\prime}\right)+D_{3}\left(4F_{1}F_{3}+F_{3}^{\prime}\right)+3D_{6}F_{3}^{2},\\
G_{3}= & 2\left(D_{1}F_{1}+D_{3}F_{3}\right)=2G_{1},\\
G_{4}= & D_{4}+D_{5},\\
G_{5}= & D_{2}F_{2},\\
G_{6}= & D_{2}\left(F_{2}^{2}+F_{2}^{\prime}\right),\\
G_{7}= & 2D_{2}F_{2}=2G_{5},\\
G_{8}= & D_{1}^{\prime}F_{1}+\frac{1}{2}D_{3}^{\prime}F_{3},\\
G_{9}= & D_{2}^{\prime}F_{2},\\
G_{10}= & D_{1}^{\prime}\left(F_{1}^{2}+F_{1}^{\prime}\right)+D_{3}^{\prime}\left(\frac{1}{2}F_{3}^{\prime}+2F_{1}F_{3}\right)+D_{6}^{\prime}F_{3}^{2},\\
G_{11}= & D_{2}^{\prime}\left(F_{2}^{2}+F_{2}^{\prime}\right),\\
G_{12}= & 2D_{1}^{\prime}F_{1}+D_{3}^{\prime}F_{3},\\
G_{13}= & 2D_{2}^{\prime}F_{2}=2G_{9},\\
G_{14}= & D_{4}^{\prime}+D_{5}^{\prime}.
\end{align}
The terms $D_{1}r_{\rho}$, $D_{2}\lambda^{2}r_{z}$, $D_{1}^{\prime}r_{\rho}^{2}$,
and $D_{2}^{\prime}r_{z}^{2}$ come from the trapping term $L_{pot}$,
which can be neglected in the case of a freely expanding condensate.
The parameter $\gamma$ indicates the terms generated by the atomic
interaction potential, while the fractions proportional to $r_{\rho}^{-2}$
and $r_{\rho}^{-3}$ come from the kinetic energy contribution due
to the presence of the vortex with charge $\ell$. The remaining factors
represent the coupling between the outer dimensions of the condensate
and the vortex core.

Making the velocities ($\dot{r_{\rho}}$, $\dot{r_{z}}$ ,$\dot{\alpha}$)
and accelerations ($\ddot{r_{\rho}}$, $\ddot{r_{z}}$, $\ddot{\alpha}$)
equal to zero leads to the equations for the stationary solution:
\begin{align}
D_{1}r_{\rho0} & =G_{4}\frac{\ell^{2}}{r_{\rho0}^{3}}+D_{7}\frac{4\pi\gamma}{r_{\rho0}^{3}r_{z0}},\label{eq:ss1}\\
D_{2}\lambda^{2}r_{z0} & =D_{7}\frac{2\pi\gamma}{r_{\rho0}^{2}r_{z0}^{2}},\label{eq:ss2}\\
D_{1}^{\prime}r_{\rho0}^{2}+D_{2}^{\prime}\lambda^{2}r_{z0}^{2} & =-G_{14}\frac{\ell^{2}}{r_{\rho0}^{2}}-D_{7}^{\prime}\frac{4\pi\gamma}{r_{\rho0}^{2}r_{z0}},\label{eq:ss3}
\end{align}
where $r_{\rho}$, $r_{z}$, and $r_{\xi}$ take their respective
equilibrium values $r_{\rho0}$, $r_{z0}$, and $r_{\xi0}$. We apply
the Newton's method to solve the coupled stationary equations (\ref{eq:ss1})--(\ref{eq:ss3}).
The value of the atomic interaction parameter used from now on in
this paper is $\gamma=800$, which is close to the value used in Rubidium
experiments \cite{emanuel}.

\section{Collective excitations}

{\indent}

For small deviations from the equilibrium configuration, we assume
$r_{\rho}\left(t\right)\rightarrow r_{\rho0}+\delta\rho\left(t\right)$,
$r_{z}\left(t\right)\rightarrow r_{z0}+\delta z\left(t\right)$, $\alpha\left(t\right)\rightarrow\alpha_{0}+\delta\alpha\left(t\right)$,
and neglect all terms of order two or higher in (\ref{eq:2d})--(\ref{eq:2f}).
This leads to the linearized matrix equation
\begin{eqnarray}
 &  & \!\!\begin{pmatrix}D_{1} & 0 & G_{1}r_{\rho0}\\
0 & D_{2} & G_{5}r_{z0}\\
D_{1}^{\prime}r_{\rho0} & D_{2}^{\prime}r_{z0} & G_{8}r_{\rho0}^{2}\!\!+\!\! G_{9}r_{z0}^{2}
\end{pmatrix}\!\!\!\!\begin{pmatrix}\ddot{\delta\rho}\\
\ddot{\delta z}\\
\ddot{\delta\alpha}
\end{pmatrix}+\nonumber \\
 &  & \!\!\begin{pmatrix}D_{1}\!\!+\!\!3G_{4}\frac{\ell^{2}}{r_{\rho0}^{4}}\!\!+\!\! D_{7}\frac{12\pi\gamma}{r_{\rho0}^{4}r_{z0}} & D_{7}\frac{4\pi\gamma}{r_{\rho0}^{3}r_{z0}^{2}} & D_{1}^{\prime}r_{\rho0}\!\!-\!\! G_{4}^{\prime}\frac{\ell^{2}}{r_{\rho0}^{3}}\!\!-\!\! D_{7}^{\prime}\frac{4\pi\gamma}{r_{\rho0}^{3}r_{z0}}\\
D_{7}\frac{4\pi\gamma}{r_{\rho0}^{3}r_{z0}^{2}} & D_{2}\lambda^{2}\!\!+\!\! D_{7}\frac{4\pi\gamma}{r_{\rho0}^{2}r_{z0}^{3}} & D_{2}^{\prime}\lambda^{2}r_{z0}\!\!-\!\! D_{7}^{\prime}\frac{2\pi\gamma}{r_{\rho0}^{2}r_{z0}^{2}}\\
2D_{1}^{\prime}r_{\rho0}\!\!-\!\!2G_{14}\frac{\ell^{2}}{r_{\rho0}^{3}}\!\!-\!\! D_{7}^{\prime}\frac{8\pi\gamma}{r_{\rho0}^{3}r_{z0}} & 2D_{2}^{\prime}\lambda^{2}r_{z0}\!\!-\!\! D_{7}^{\prime}\frac{4\pi\gamma}{r_{\rho0}^{2}r_{z0}^{2}} & D_{1}^{\prime\prime}r_{\rho0}^{2}\!\!+\!\! D_{2}^{\prime\prime}\lambda^{2}r_{z0}^{2}\!\!+\!\! G_{14}^{\prime}\frac{\ell^{2}}{r_{\rho0}^{2}}\!\!+\!\! D_{7}^{\prime\prime}\frac{4\pi\gamma}{r_{\rho0}^{2}r_{z0}}
\end{pmatrix}\!\!\!\!\begin{pmatrix}\delta\rho\\
\delta z\\
\delta\alpha
\end{pmatrix}\nonumber \\
 &  & =0,\label{eq:3abc}
\end{eqnarray}
which defines the matrices $M$ and $V$, appearing in Eq.(\ref{eq:2}).
Solving the characteristic equation,
\begin{equation}
\det\left(M^{-1}V-\varpi^{2}I\right)=0,\label{eq:55}
\end{equation}
results in the frequency of the collective modes of oscillation. Now
the determinants $\det M$ and $\det V$ are both positive for $\ell=1$.
Meaning that we are in the lower energy state for the case of a central
vortex in a Bose-Einstein condensate. 
\begin{figure}
\centering
\includegraphics[scale=0.6]{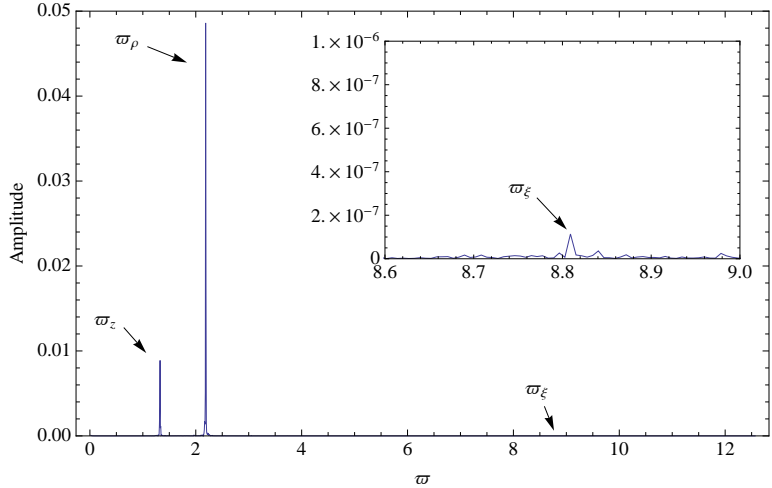}

\caption{Excitation spectrum obtained from a numerical simulation of the GPE
\cite{xmds}. We set $\gamma=800$, $\ell=1$, $\tilde{\mu}=20.74$,
and $\lambda=0.9$.$\varpi_{n}$ are the frequencies of the oscillation
modes from less energetic ($\varpi_{z}$) to more energetic ($\varpi_{\xi}$).
The analytical values are $\varpi_{z}=1.317$, $\varpi_{\rho}=2.166$,
and $\varpi_{\xi}=8.874$.}

\label{figN}
\end{figure}
\begin{figure}
\centering

\subfloat[$\ell=1$]{\includegraphics[scale=0.65]{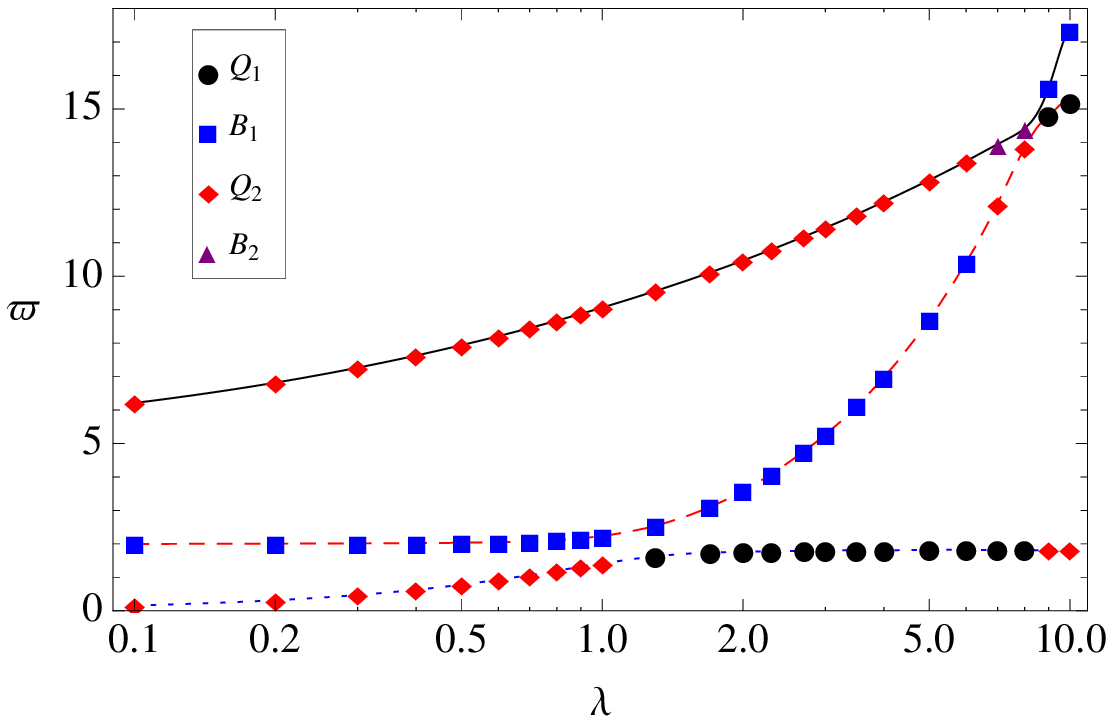}

\label{fig4a}}\subfloat[$\ell=2$]{\includegraphics[scale=0.65]{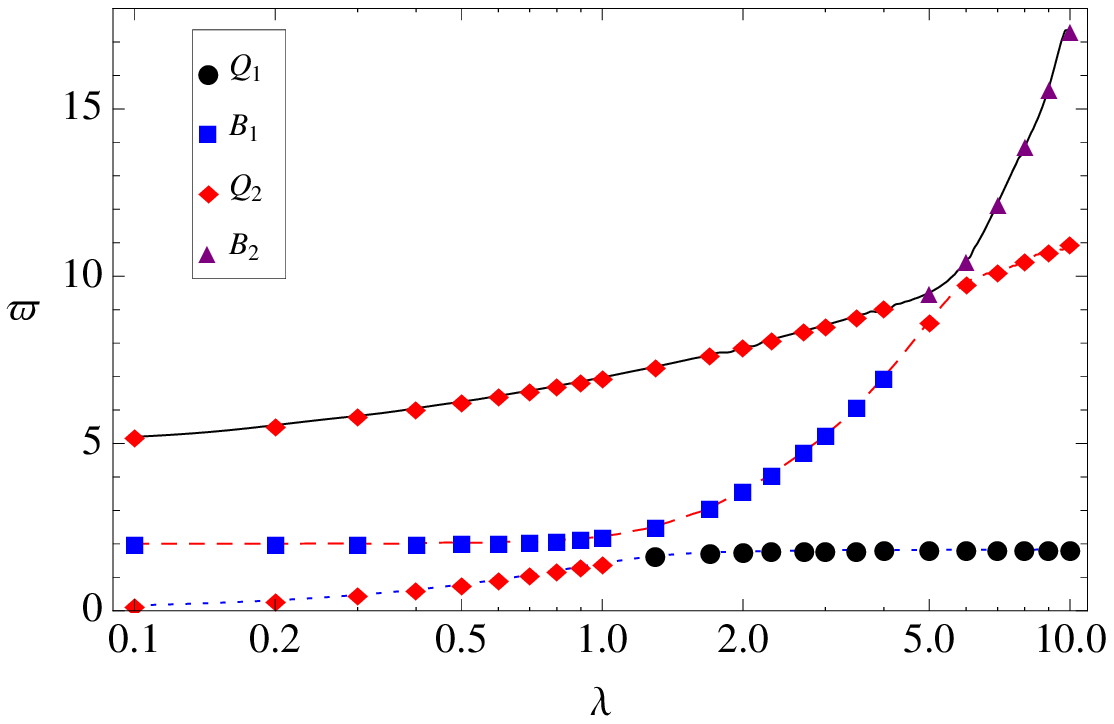}

\label{fig4b}

}

\caption{(Color online) Oscillation frequencies as a function of trap anisotropy
in a condensate containing a singly (a) and doubly (b) charged vortex
at its center. Solid (black) line is $\varpi_{\xi}$, dashed (red)
line is $\varpi_{\rho}$, and dotted (blue) line is $\varpi_{z}$.}

\label{fig4}
\end{figure}
\begin{figure}
\subfloat[$B_{1}$]{\includegraphics[scale=0.4]{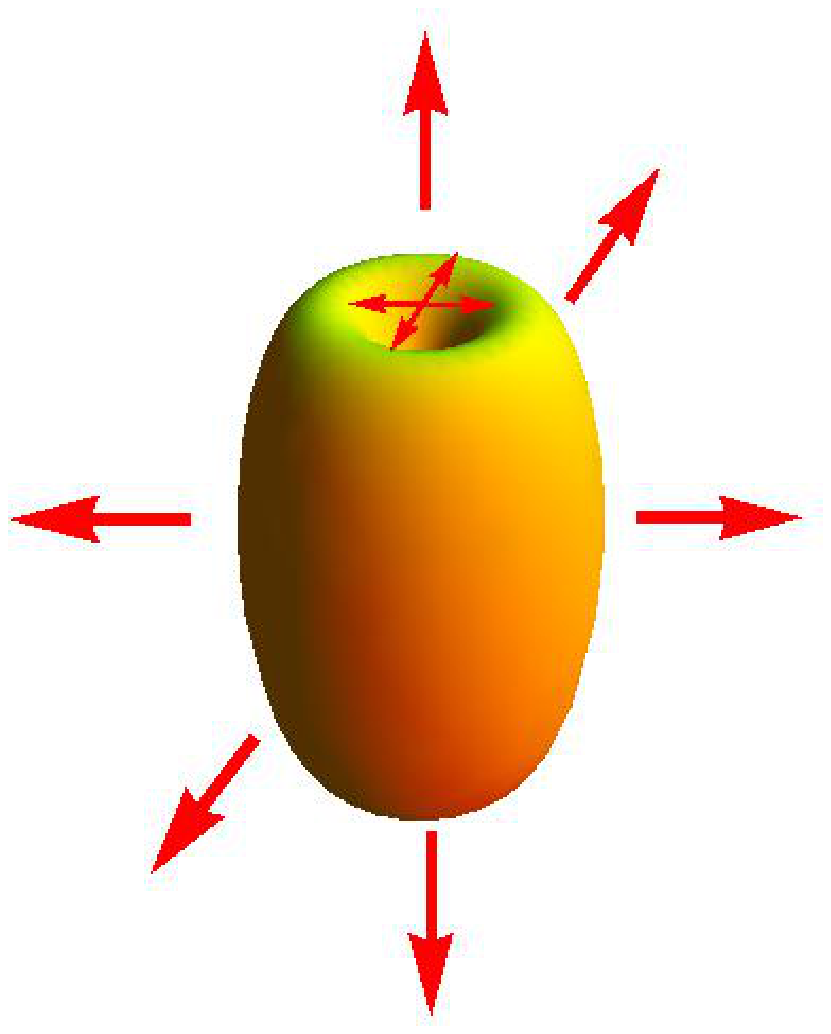}

\label{figM-1}}\subfloat[$Q_{1}$]{\includegraphics[scale=0.4]{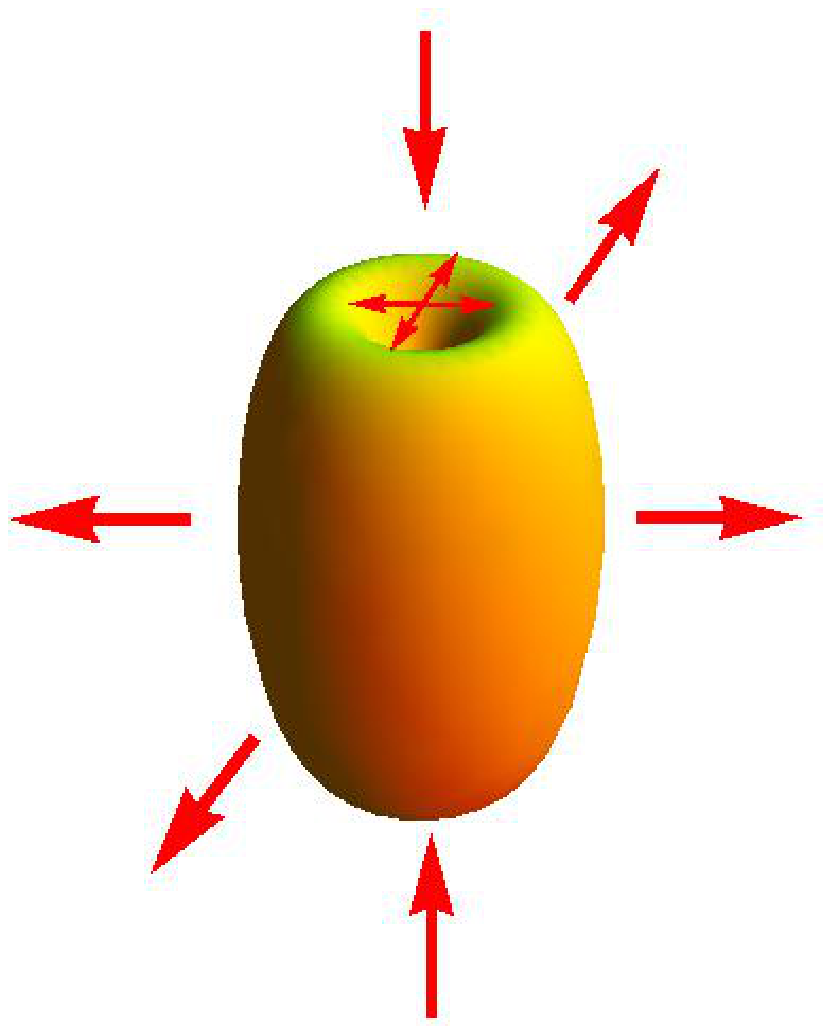}

\label{figM-2}}

\subfloat[$B_{2}$]{\includegraphics[scale=0.4]{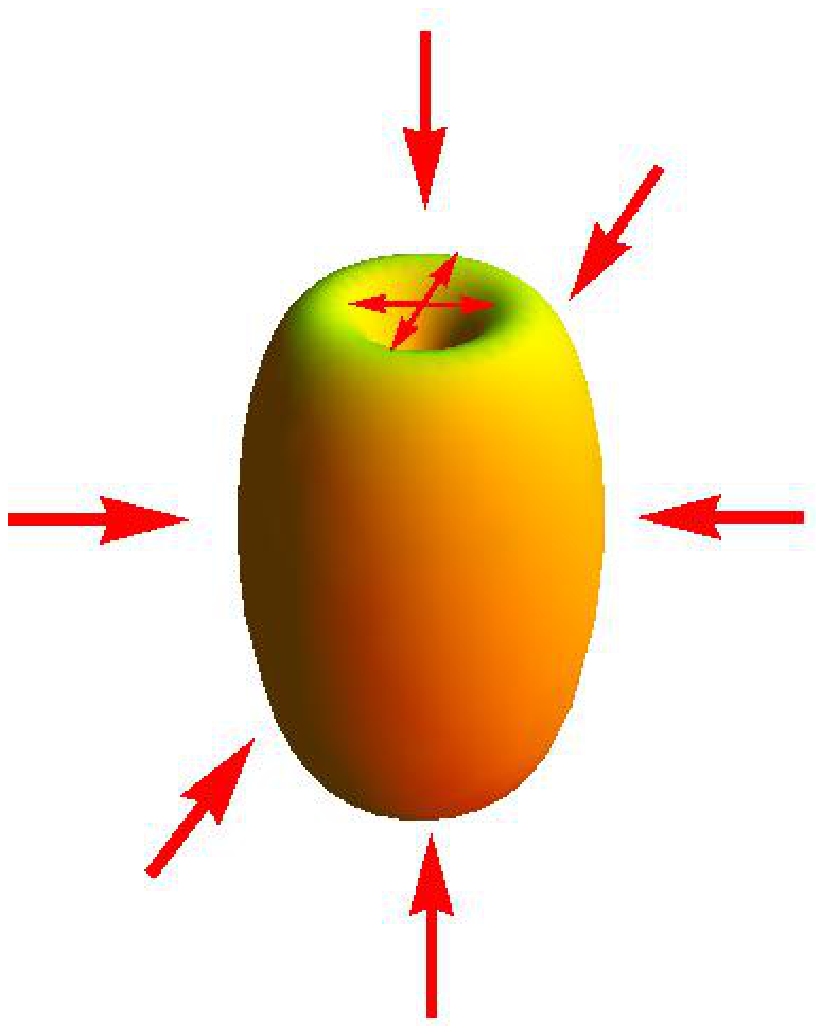}

\label{figM-3}}\subfloat[$Q_{2}$]{\includegraphics[scale=0.4]{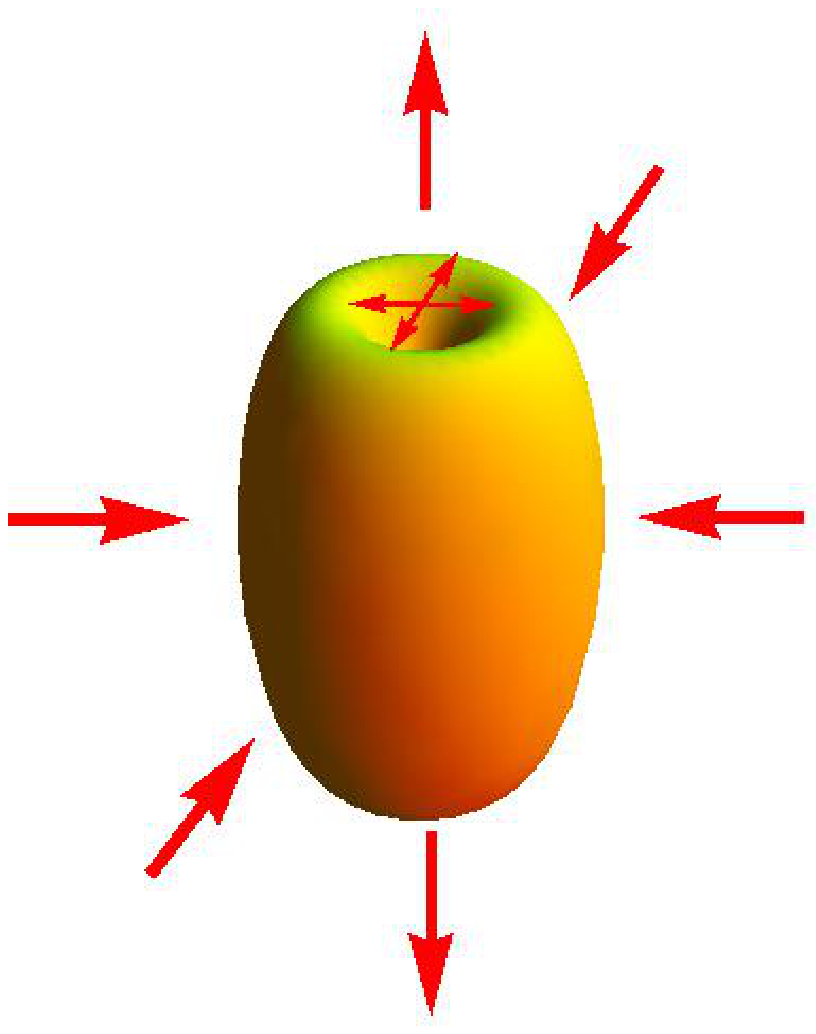}

\label{figM-4}}

\caption{(Color online) Schematic representation of collective modes. $B_{1}$
mode has all components oscillating in phase. $B_{2}$ mode has $r_{\xi}$
oscillating out of phase with $r_{\rho}$ and $r_{z}$. $Q_{1}$ mode
has $r_{z}$ oscillation out of phase with $r_{\xi}$ and $r_{\rho}$.
$Q_{2}$ mode has $r_{\rho}$ oscillation out of phase with $r_{\xi}$
and $r_{z}$.}

\label{figM}
\end{figure}
 Since (\ref{eq:55}) is a cubic equation of $\varpi^{2}$, we have
three pair of frequencies $\pm\varpi_{n}$ ($n=z,\rho,\xi$). There
are three frequencies $\varpi_{n}$ and four modes of oscillation
in total, of which only three modes can be simultaneously observed
depending on the anisotropy $\lambda$ of harmonic potential as shown
in fig.\ref{fig4}. Among these four modes, two of them represent
monopole oscillations while the other two represent quadrupole oscillations
of the atomic cloud. The $B_{1}$ mode (fig.\ref{figM-1}) is characterized
by having all condensate components $r_{i}$ ($i=z,\rho,\xi$) oscillating
in phase, however $B_{2}$ mode (fig.\ref{figM-3}) presents $r_{\xi}$
oscillating out of phase with $r_{\rho}$ and $r_{z}$. The $Q_{1}$
mode (fig.\ref{figM-2}) shows that $r_{z}$ oscillation is out of
phase with $r_{\xi}$ and $r_{\rho}$, which are in phase with each
other. However, in $Q_{2}$ mode (fig.\ref{figM-4}) the oscillations
of $r_{z}$ and $r_{\xi}$ are in phase with each other, being the
$r_{\rho}$ oscillation out of phase. Extrapolating to an ideal situation
where $\gamma=0$, the equations of motions (\ref{eq:2d})--(\ref{eq:2f})
can be decoupled. This way, the $\varpi_{z}$ (lower frequency) represents
only a $r_{z}$ oscillation, $\varpi_{\rho}$ (middle frequency) represents
only a $r_{\rho}$ oscillation, and $\varpi_{\xi}$ (upper frequency)
represents only a $r_{\xi}$ oscillation.

Numerical simulations where performed in order to validate our results
(fig.\ref{figN}). Frequency values $\varpi_{n}$ in the variational
calculations differ from numerical values by less than 1\%. 

In fig.\ref{fig4a}, for $0.1\leq\lambda\leq1$, exist two $Q_{2}$-like
modes. The difference between them comes from the fact that vortex
core oscillation amplitude which is two orders of magnitude lower
at the less energetic mode. The same happens when $\ell=2$ (fig.\ref{fig4b}),
i.e., the vortex core is almost still for the lower frequency in the
same interval of $\lambda$.

The solid lines in fig.\ref{fig4} correspond to the mode with largest
amplitude for the vortex-core oscillations. As can be seen, the excitation
frequency $\varpi_{\xi}$ of this mode lowers as the vortex circulation
increases. It means that the energy necessary to excite it will be
lower if $\ell$ is increased. However, we must point out that our
results apply only for the cases where $r_{\xi}\ll r_{\rho}$.

\section{Scattering length modulation}

{\indent}

One of the mechanisms used for exciting collective modes is via modulation
of the s-wave scattering length. This technique has been already applied
to excite the lowest-lying quadrupole mode in a Lithium experiment
\cite{cm1}. Therefore, we consider the time-dependent scattering
length:
\begin{equation}
a_{s}(t)=a_{0}+\delta a\cos\left(\Omega t\right).
\end{equation}
This is equivalent to make $\gamma\rightarrow\gamma(\tau)$, thus
giving:
\begin{equation}
\gamma(\tau)=\gamma_{0}+\delta\gamma\cos\left(\Omega\tau\right).\label{eq:slm}
\end{equation}
Where $\gamma_{0}$ is the average value of the interaction parameter
$\gamma\left(\tau\right)$, $\delta\gamma$ is the modulation amplitude,
and $\Omega$ is the excitation frequency. Substituting (\ref{eq:slm})
into (\ref{eq:3abc}) and keeping only first-order terms ($\delta\rho$,
$\delta z$, $\delta\alpha$, and $\delta\gamma$), we obtain a nonhomogeneous
linear equation

\begin{equation}
M\ddot{\delta}+V\delta=P\cos\left(\Omega\tau\right)\label{eq:modula=0000E7=0000E3o}
\end{equation}

with
\begin{equation}
P=2\pi\delta\gamma\begin{pmatrix}\frac{2D_{7}}{r_{\rho0}^{3}r_{z0}}\\
\frac{D_{7}}{r_{\rho0}^{2}r_{z0}^{2}}\\
\frac{D_{7}^{\prime}}{r_{\rho0}^{2}r_{z0}}
\end{pmatrix}.
\end{equation}

A particular solution of (\ref{eq:modula=0000E7=0000E3o}) is

\begin{equation}
\delta_{\gamma}\left(\tau\right)=\left(M^{-1}V-\Omega^{2}\right)^{-1}M^{-1}P\cos\left(\Omega\tau\right).
\end{equation}

Projecting the vector $\delta_{\gamma}\left(\tau\right)$ in the base
$\delta_{n}$ ($n=z,\rho,\xi$) of the eigenvectors of the homogenous
equation associated to Eq.(\ref{eq:modula=0000E7=0000E3o}), we obtain

\begin{equation}
\left\langle \delta_{n}|\delta_{\gamma}\left(\tau\right)\right\rangle =\frac{\left\langle \delta_{n}|M^{-1}P\right\rangle }{\varpi_{n}^{2}-\Omega^{2}}\cos\left(\Omega\tau\right).\label{eq:ngama}
\end{equation}

\begin{figure}
\centering

\includegraphics[scale=0.8]{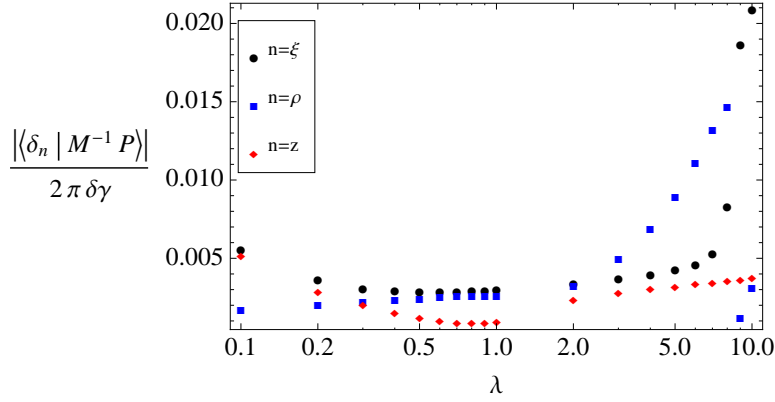}

\caption{(Color online) Overlap between $\left|M^{-1}P\right\rangle $ and
normal modes $\left|\delta_{n}\right\rangle $ as a function of the
trap anisotropy with circulation $\ell=1$. The scalar product $\left|\left\langle \delta_{n}|M^{-1}P\right\rangle \right|$
is always positive.}

\label{fig nMP}
\end{figure}

\begin{figure}
\centering
\subfloat[]{\includegraphics[scale=0.47]{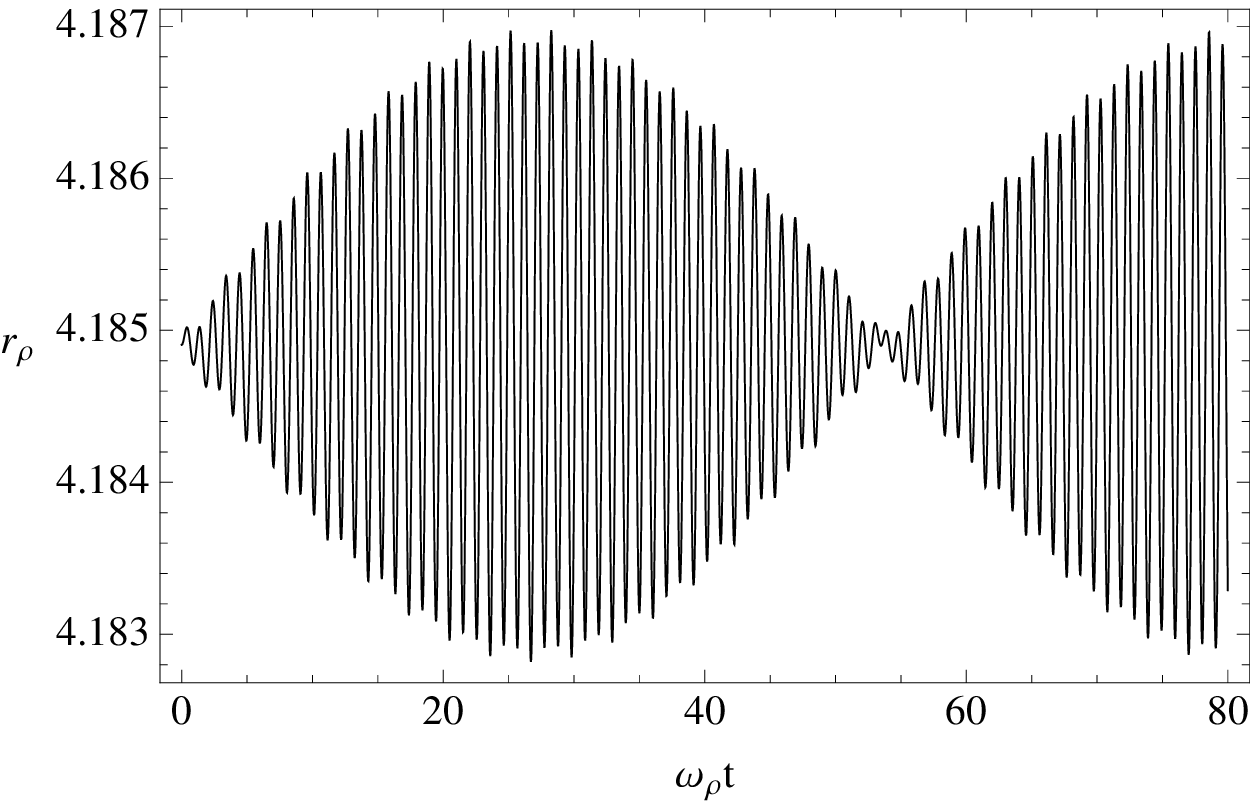}

}\subfloat[]{\includegraphics[scale=0.47]{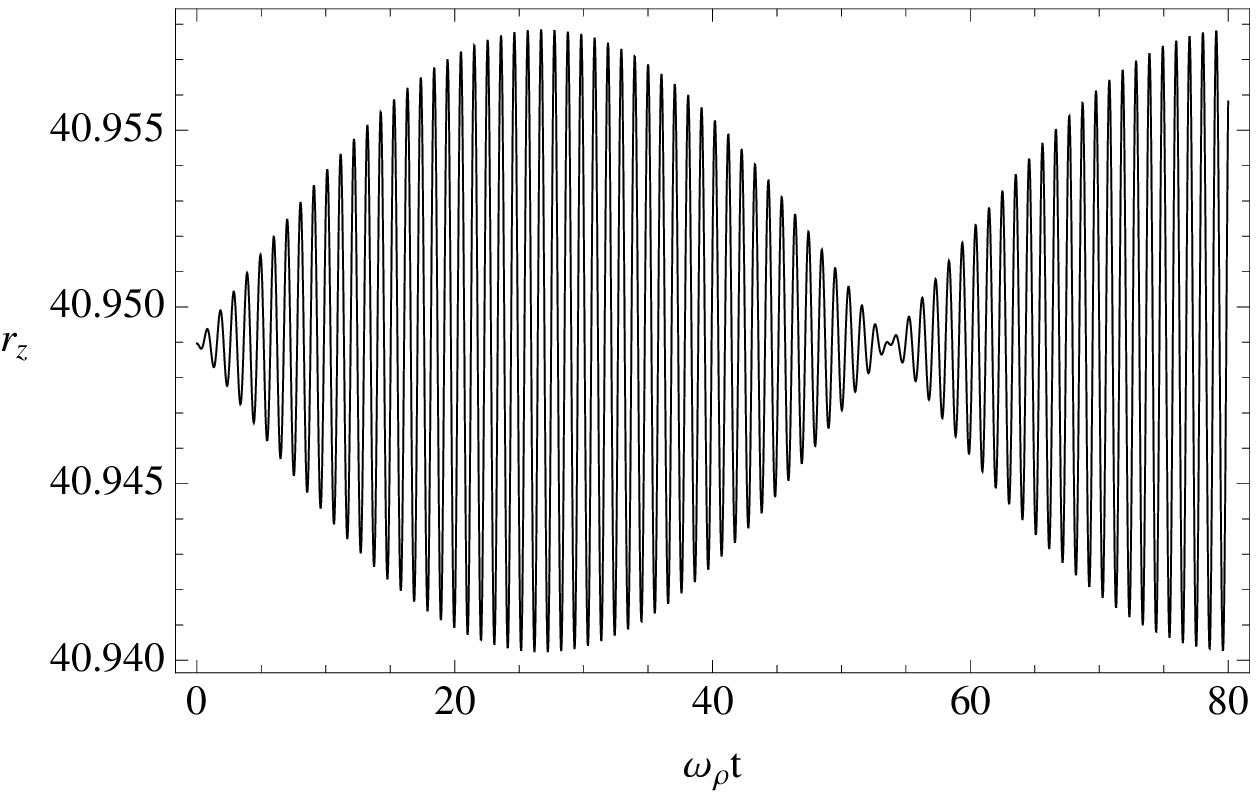}

}

\subfloat[]{\includegraphics[scale=0.47]{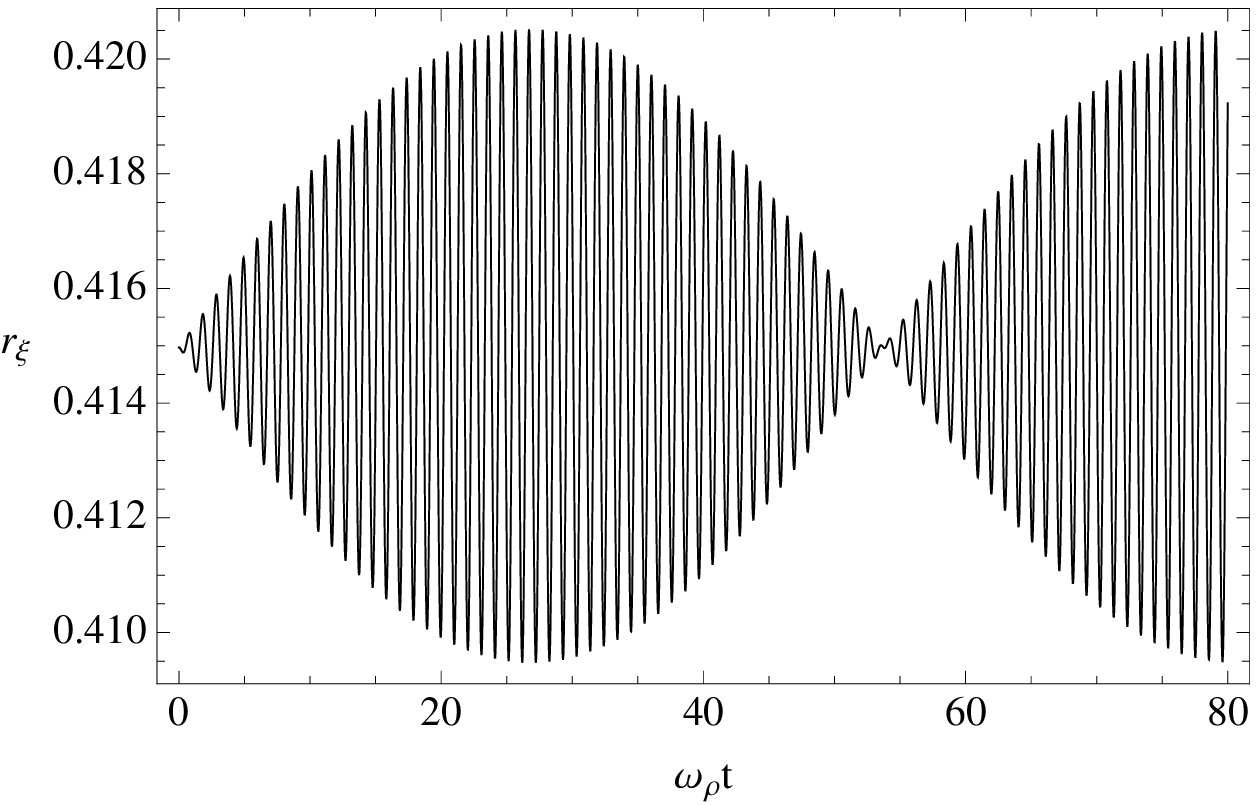}

}\subfloat[]{\includegraphics[scale=0.5]{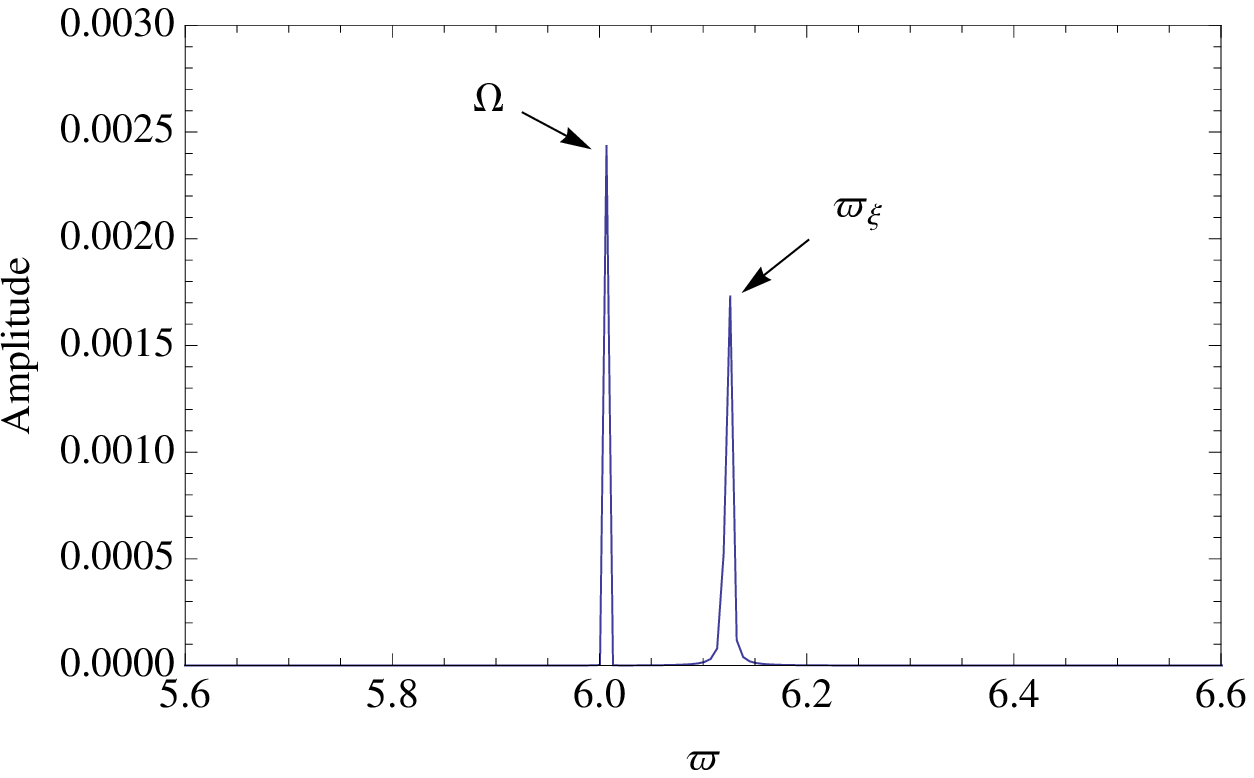}

}

\caption{Numerical solution of Eqs.(\ref{eq:2d})--(\ref{eq:2f}) with a time-dependent
interaction $\gamma\left(\tau\right)$ (a, b, and c). (d) is the excitation
spectrum obtained from variational calculation, where $\varpi_{\xi}\approx6.13$
is close to the value calculated in Eq.(\ref{eq:55}). We excited
the collective mode $Q_{2}$ ($\varpi_{\xi}=6.21$) of a condensate
with cigar shape ($\lambda=0.1$, $\gamma_{0}=800$) via scattering
modulation with amplitude $\delta\gamma=0.4$ and frequency $\Omega=6$. }

\label{fig5}

\end{figure}

Since $\left|\left\langle \delta_{n}|M^{-1}P\right\rangle \right|>0$
(fig.\ref{fig nMP}), it shows that specific collective modes can
be excited using a scattering length modulation with small amplitude
$\delta\gamma$ and frequency $\Omega$ close to one of the resonance
frequency $\varpi_{n}$. In fig.\ref{fig5}, we see the results from
a numerical solution of Eqs.(\ref{eq:2d})--(\ref{eq:2f}) considering
a time-dependent interaction $\gamma\left(\tau\right)$ according
to Eq.(\ref{eq:slm}). There we can see the beat behavior corresponding
to a superposition of the frequencies $\Omega=6$ and $\varpi_{\xi}=6.13$.


\section{Free expansion}

{\indent}
\begin{figure}
\centering
\subfloat[$\ell=1$]{\includegraphics[scale=0.65]{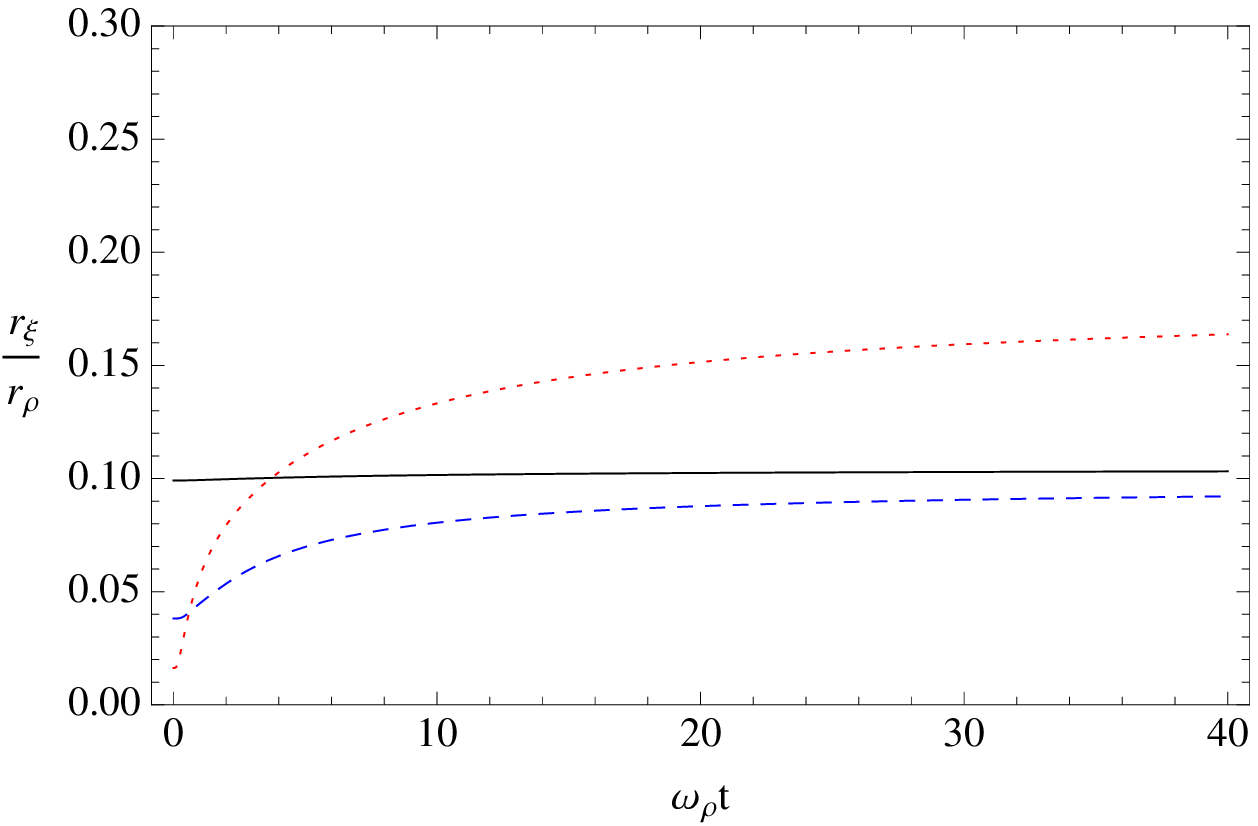}

\label{fig7a}}\subfloat[$\ell=2$]{\includegraphics[scale=0.65]{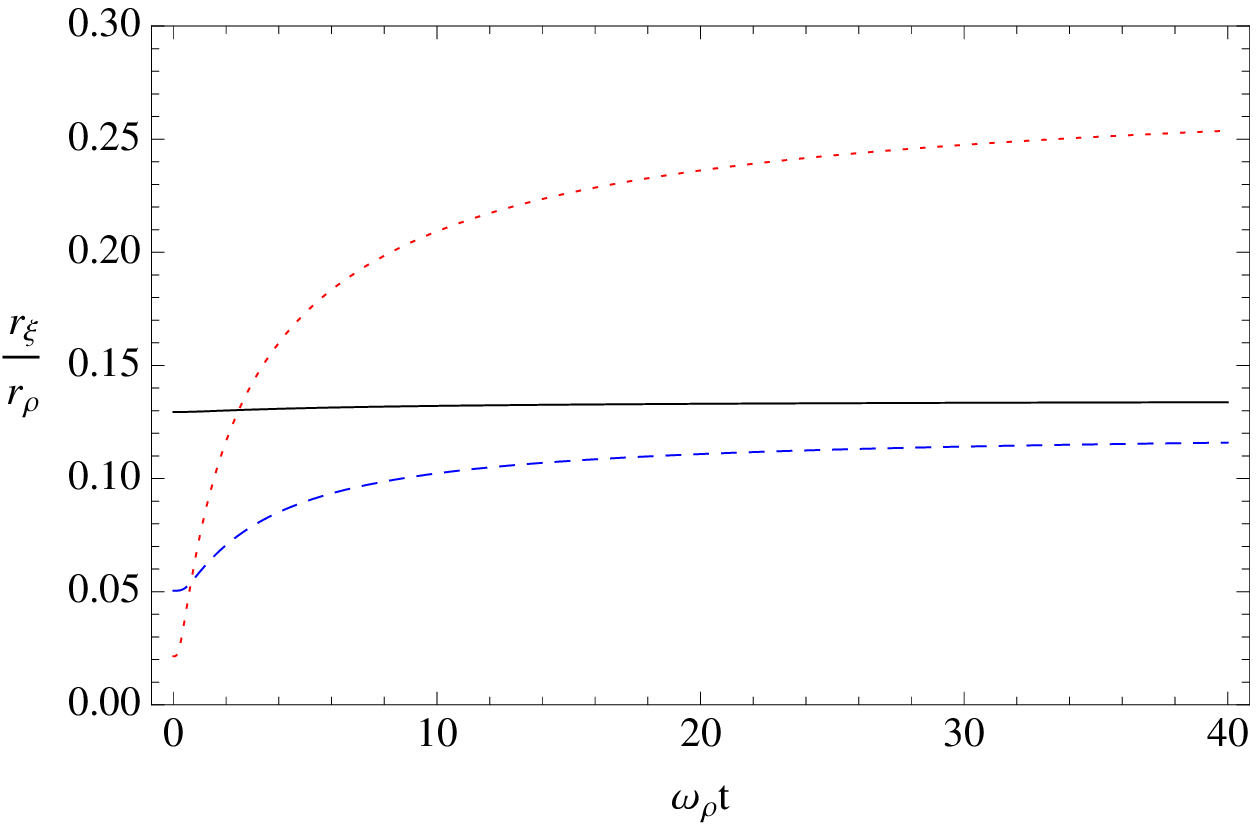}

\label{fig7b}

}

\caption{(Color online) Ratio between vortex core size and radial cloud size
for different trap anisotropies while in free expansion. The solid
(black) line corresponds to a prolate condensate ($\lambda=0.1$),
the dashed (blue) line to the isotropic case ($\lambda=1$), and the
dotted (red) line to a oblate condensate ($\lambda=8$).}

\label{fig7}
\end{figure}

The time-of-flight pictures constitute the most common method to measure
vortices in BEC. This method of switch off the magneto-optical trap
and letting atomic cloud expand freely for some time, typically ten
milliseconds and then taking a picture of the expanded cloud \cite{rev3,MIT,obv,turb,vor1,vp}.
For this purpose we use the equations of motion (\ref{eq:2d})--(\ref{eq:2f})
without the terms arising from the harmonic potential, i.e.,
\begin{align}
D_{1}\ddot{r_{\rho}}+G_{1}r_{\rho}\ddot{\alpha}+G_{2}r_{\rho}\dot{\alpha}^{2}+G_{3}\dot{r_{\rho}}\dot{\alpha}-\frac{G_{4}}{r_{\rho}^{3}}-\frac{4D_{7}\gamma}{r_{\rho}^{3}r_{z}} & =0,\\
D_{2}\ddot{r_{z}}+G_{5}r_{z}\ddot{\alpha}+G_{6}r_{z}\dot{\alpha}^{2}+G_{7}\dot{r_{z}}\dot{\alpha}-\frac{2D_{7}\gamma}{r_{\rho}^{2}r_{z}^{2}} & =0,\\
D_{1}^{\prime}r_{\rho}\ddot{r_{\rho}}+D_{2}^{\prime}r_{z}\ddot{r_{z}}+\left(G_{8}r_{\rho}^{2}+G_{9}r_{z}^{2}\right)\ddot{\alpha}+\left(G_{10}r_{\rho}^{2}+G_{11}r_{z}^{2}\right)\dot{\alpha}^{2}\nonumber \\
+\left(G_{12}r_{\rho}\dot{r_{\rho}}+G_{13}r_{z}\dot{r_{z}}\right)\dot{\alpha}+\frac{G_{14}}{r_{\rho}^{3}}+\frac{4D_{7}^{\prime}\gamma}{r_{\rho}^{2}r_{z}} & =0,
\end{align}
whose initial conditions are given by the stationary equations (\ref{eq:ss1})--(\ref{eq:ss3}).
This result agrees with our preview work \cite{rpteles1}, where the
free expansion of the vortex core is given in fig.\ref{fig7}. In
\cite{rpteles1}, the fig.\ref{fig7b} could not be calculated since
the authors considered the healing length as an approximation to the
vortex core radius which is only valid for $\ell=1$.

\section{Conclusions}

{\indent}

In this paper we proposed a modification in the wave function phase
commonly used with the variational method which corrects the imaginary
frequencies of collective modes when we have a parameter describing
non-physical vortex core dynamics with $\ell=1$.

Here we consider variational phase parameters corresponding to each
parameter in wave function amplitude, respectively. This way, we were
able to describe the dynamics of both vortex core and the external
dimensions of the condensate which agrees with the numerical simulations
of the GPE. Although we observe four modes of oscillation in total,
only three of them can be simultaneously observed depending on the
trap anisotropy.

We also demonstrate that these oscillation modes can be excited by
modulating the s-wave scattering length using the same experimental
techniques as in Ref. \cite{cm1}.

Finally, we analyzed the time-of-flight dynamics of the vortex core
with different circulations in order to complement the results in
Ref. \cite{rpteles1}.
\begin{acknowledgments}
We acknowledge the financial support of from the National Council
for the Improvement of Higher Education (CAPES) and from the State
of S\~ao Paulo Foundation for Research Support (FAPESP).
\end{acknowledgments}
\pagebreak{}

\bibliography{refdoc}

\begin{thebibliography}{28}
\expandafter\ifx\csname natexlab\endcsname\relax\def\natexlab#1{#1}\fi
\expandafter\ifx\csname bibnamefont\endcsname\relax
  \def\bibnamefont#1{#1}\fi
\expandafter\ifx\csname bibfnamefont\endcsname\relax
  \def\bibfnamefont#1{#1}\fi
\expandafter\ifx\csname citenamefont\endcsname\relax
  \def\citenamefont#1{#1}\fi
\expandafter\ifx\csname url\endcsname\relax
  \def\url#1{\texttt{#1}}\fi
\expandafter\ifx\csname urlprefix\endcsname\relax\def\urlprefix{URL }\fi
\providecommand{\bibinfo}[2]{#2}
\providecommand{\eprint}[2][]{\url{#2}}

\bibitem[{\citenamefont{Pollack et~al.}(2010)\citenamefont{Pollack, Dries,
  Hulet, Magalh{\~a}es, Henn, Ramos, Caracanhas, and Bagnato}}]{cm1}
\bibinfo{author}{\bibfnamefont{S.~E.} \bibnamefont{Pollack}},
  \bibinfo{author}{\bibfnamefont{D.}~\bibnamefont{Dries}},
  \bibinfo{author}{\bibfnamefont{R.~G.} \bibnamefont{Hulet}},
  \bibinfo{author}{\bibfnamefont{K.~M.~F.} \bibnamefont{Magalh{\~a}es}},
  \bibinfo{author}{\bibfnamefont{E.~A.~L.} \bibnamefont{Henn}},
  \bibinfo{author}{\bibfnamefont{E.~R.~F.} \bibnamefont{Ramos}},
  \bibinfo{author}{\bibfnamefont{M.~A.} \bibnamefont{Caracanhas}},
  \bibnamefont{and} \bibinfo{author}{\bibfnamefont{V.~S.}
  \bibnamefont{Bagnato}}, \bibinfo{journal}{Physical Review A}
  \textbf{\bibinfo{volume}{81}}, \bibinfo{pages}{053627}
  (\bibinfo{year}{2010}).

\bibitem[{\citenamefont{Stringari}(1996)}]{c-ex1}
\bibinfo{author}{\bibfnamefont{S.}~\bibnamefont{Stringari}},
  \bibinfo{journal}{Physical Review Letters} \textbf{\bibinfo{volume}{77}},
  \bibinfo{pages}{2360} (\bibinfo{year}{1996}).

\bibitem[{\citenamefont{P{\'e}rez-Garc{\'\i}a
  et~al.}(1996)\citenamefont{P{\'e}rez-Garc{\'\i}a, Michinel, Cirac,
  Lewenstein, and Zoller}}]{perez2}
\bibinfo{author}{\bibfnamefont{V.~M.} \bibnamefont{P{\'e}rez-Garc{\'\i}a}},
  \bibinfo{author}{\bibfnamefont{H.}~\bibnamefont{Michinel}},
  \bibinfo{author}{\bibfnamefont{J.~I.} \bibnamefont{Cirac}},
  \bibinfo{author}{\bibfnamefont{M.}~\bibnamefont{Lewenstein}},
  \bibnamefont{and} \bibinfo{author}{\bibfnamefont{P.}~\bibnamefont{Zoller}},
  \bibinfo{journal}{Physical Review Letters} \textbf{\bibinfo{volume}{77}},
  \bibinfo{pages}{5320} (\bibinfo{year}{1996}).

\bibitem[{\citenamefont{Dalfovo et~al.}(1999)\citenamefont{Dalfovo, Giorgini,
  Pitaevskii, and Stringari}}]{rev1}
\bibinfo{author}{\bibfnamefont{F.}~\bibnamefont{Dalfovo}},
  \bibinfo{author}{\bibfnamefont{S.}~\bibnamefont{Giorgini}},
  \bibinfo{author}{\bibfnamefont{L.~P.} \bibnamefont{Pitaevskii}},
  \bibnamefont{and}
  \bibinfo{author}{\bibfnamefont{S.}~\bibnamefont{Stringari}},
  \bibinfo{journal}{Reviews of Modern Physics} \textbf{\bibinfo{volume}{71}},
  \bibinfo{pages}{463} (\bibinfo{year}{1999}).

\bibitem[{\citenamefont{Courteille et~al.}(2001)\citenamefont{Courteille,
  Bagnato, and Yukalov}}]{rev3}
\bibinfo{author}{\bibfnamefont{P.~W.} \bibnamefont{Courteille}},
  \bibinfo{author}{\bibfnamefont{V.~S.} \bibnamefont{Bagnato}},
  \bibnamefont{and} \bibinfo{author}{\bibfnamefont{V.~I.}
  \bibnamefont{Yukalov}}, \bibinfo{journal}{Laser Physics}
  \textbf{\bibinfo{volume}{11}}, \bibinfo{pages}{659} (\bibinfo{year}{2001}).

\bibitem[{\citenamefont{Busch et~al.}(1997)\citenamefont{Busch, Cirac,
  P{\'e}rez-Garc{\'\i}a, and Zoller}}]{2-comp}
\bibinfo{author}{\bibfnamefont{T.}~\bibnamefont{Busch}},
  \bibinfo{author}{\bibfnamefont{J.~I.} \bibnamefont{Cirac}},
  \bibinfo{author}{\bibfnamefont{V.~M.} \bibnamefont{P{\'e}rez-Garc{\'\i}a}},
  \bibnamefont{and} \bibinfo{author}{\bibfnamefont{P.}~\bibnamefont{Zoller}},
  \bibinfo{journal}{Physical Review A} \textbf{\bibinfo{volume}{56}},
  \bibinfo{pages}{2978} (\bibinfo{year}{1997}).

\bibitem[{\citenamefont{Zhang and Liu}(2011)}]{cm2}
\bibinfo{author}{\bibfnamefont{Z.}~\bibnamefont{Zhang}} \bibnamefont{and}
  \bibinfo{author}{\bibfnamefont{W.~V.} \bibnamefont{Liu}},
  \bibinfo{journal}{Physical Review A} \textbf{\bibinfo{volume}{83}},
  \bibinfo{pages}{023617} (\bibinfo{year}{2011}).

\bibitem[{\citenamefont{Heiselberg}(2004)}]{cm3}
\bibinfo{author}{\bibfnamefont{H.}~\bibnamefont{Heiselberg}},
  \bibinfo{journal}{Physical Review Letters} \textbf{\bibinfo{volume}{93}},
  \bibinfo{pages}{040402} (\bibinfo{year}{2004}).

\bibitem[{\citenamefont{Altmeyer et~al.}(2007)\citenamefont{Altmeyer, Riedl,
  Kohstall, Wright, Geursen, Bartenstein, Chin, Denschlag, and Grimm}}]{cm5}
\bibinfo{author}{\bibfnamefont{A.}~\bibnamefont{Altmeyer}},
  \bibinfo{author}{\bibfnamefont{S.}~\bibnamefont{Riedl}},
  \bibinfo{author}{\bibfnamefont{C.}~\bibnamefont{Kohstall}},
  \bibinfo{author}{\bibfnamefont{M.~J.} \bibnamefont{Wright}},
  \bibinfo{author}{\bibfnamefont{R.}~\bibnamefont{Geursen}},
  \bibinfo{author}{\bibfnamefont{M.}~\bibnamefont{Bartenstein}},
  \bibinfo{author}{\bibfnamefont{C.}~\bibnamefont{Chin}},
  \bibinfo{author}{\bibfnamefont{J.~H.} \bibnamefont{Denschlag}},
  \bibnamefont{and} \bibinfo{author}{\bibfnamefont{R.}~\bibnamefont{Grimm}},
  \bibinfo{journal}{Physical Review Letters} \textbf{\bibinfo{volume}{98}},
  \bibinfo{pages}{040401} (\bibinfo{year}{2007}).

\bibitem[{\citenamefont{{\v C}love{\v c}ko et~al.}(2008)\citenamefont{{\v
  C}love{\v c}ko, Ga{\v z}o, Kupka, and Skyba}}]{cm4}
\bibinfo{author}{\bibfnamefont{M.}~\bibnamefont{{\v C}love{\v c}ko}},
  \bibinfo{author}{\bibfnamefont{E.}~\bibnamefont{Ga{\v z}o}},
  \bibinfo{author}{\bibfnamefont{M.}~\bibnamefont{Kupka}}, \bibnamefont{and}
  \bibinfo{author}{\bibfnamefont{P.}~\bibnamefont{Skyba}},
  \bibinfo{journal}{Physical Review Letters} \textbf{\bibinfo{volume}{100}},
  \bibinfo{pages}{155301} (\bibinfo{year}{2008}).

\bibitem[{\citenamefont{Pethick and Smith}(2008)}]{pethick}
\bibinfo{author}{\bibfnamefont{C.~J.} \bibnamefont{Pethick}} \bibnamefont{and}
  \bibinfo{author}{\bibfnamefont{H.}~\bibnamefont{Smith}},
  \emph{\bibinfo{title}{Bose-einstein condensation in dilute gases}}
  (\bibinfo{publisher}{Cambridge University Press},
  \bibinfo{address}{Cambridge}, \bibinfo{year}{2008}), \bibinfo{edition}{2nd}
  ed.

\bibitem[{\citenamefont{Pitaevskii and Stringari}(2003)}]{pit-str}
\bibinfo{author}{\bibfnamefont{L.~P.} \bibnamefont{Pitaevskii}}
  \bibnamefont{and}
  \bibinfo{author}{\bibfnamefont{S.}~\bibnamefont{Stringari}},
  \emph{\bibinfo{title}{Bose-Einstein Condensation}}
  (\bibinfo{publisher}{Oxford University Press Inc}, \bibinfo{year}{2003}),
  \bibinfo{edition}{first edition} ed.

\bibitem[{\citenamefont{Svidzinsky and Fetter}(2000{\natexlab{a}})}]{fetter1}
\bibinfo{author}{\bibfnamefont{A.~A.} \bibnamefont{Svidzinsky}}
  \bibnamefont{and} \bibinfo{author}{\bibfnamefont{A.~L.}
  \bibnamefont{Fetter}}, \bibinfo{journal}{Physical Review A}
  \textbf{\bibinfo{volume}{62}}, \bibinfo{pages}{063617}
  (\bibinfo{year}{2000}{\natexlab{a}}).

\bibitem[{\citenamefont{Svidzinsky and Fetter}(2000{\natexlab{b}})}]{vi2}
\bibinfo{author}{\bibfnamefont{A.~A.} \bibnamefont{Svidzinsky}}
  \bibnamefont{and} \bibinfo{author}{\bibfnamefont{A.~L.}
  \bibnamefont{Fetter}}, \bibinfo{journal}{Physical Review Letters}
  \textbf{\bibinfo{volume}{84}}, \bibinfo{pages}{5919}
  (\bibinfo{year}{2000}{\natexlab{b}}).

\bibitem[{\citenamefont{Linn and Fetter}(2000)}]{sa-nomalmodes}
\bibinfo{author}{\bibfnamefont{M.}~\bibnamefont{Linn}} \bibnamefont{and}
  \bibinfo{author}{\bibfnamefont{A.~L.} \bibnamefont{Fetter}},
  \bibinfo{journal}{Physical Review A} \textbf{\bibinfo{volume}{61}},
  \bibinfo{pages}{063603} (\bibinfo{year}{2000}).

\bibitem[{\citenamefont{P{\'e}rez-Garc{\'\i}a and
  Garc{\'\i}a-Ripoll}(2000)}]{2-compvor}
\bibinfo{author}{\bibfnamefont{V.~M.} \bibnamefont{P{\'e}rez-Garc{\'\i}a}}
  \bibnamefont{and} \bibinfo{author}{\bibfnamefont{J.~J.}
  \bibnamefont{Garc{\'\i}a-Ripoll}}, \bibinfo{journal}{Physical Review A}
  \textbf{\bibinfo{volume}{62}}, \bibinfo{pages}{033601}
  (\bibinfo{year}{2000}).

\bibitem[{\citenamefont{P{\'e}rez-Garc{\'\i}a
  et~al.}(1997)\citenamefont{P{\'e}rez-Garc{\'\i}a, Michinel, Cirac,
  Lewenstein, and Zoller}}]{perez1}
\bibinfo{author}{\bibfnamefont{V.~M.} \bibnamefont{P{\'e}rez-Garc{\'\i}a}},
  \bibinfo{author}{\bibfnamefont{H.}~\bibnamefont{Michinel}},
  \bibinfo{author}{\bibfnamefont{J.~I.} \bibnamefont{Cirac}},
  \bibinfo{author}{\bibfnamefont{M.}~\bibnamefont{Lewenstein}},
  \bibnamefont{and} \bibinfo{author}{\bibfnamefont{P.}~\bibnamefont{Zoller}},
  \bibinfo{journal}{Physical Review A} \textbf{\bibinfo{volume}{56}},
  \bibinfo{pages}{1424} (\bibinfo{year}{1997}).

\bibitem[{\citenamefont{Svidzinsky and Fetter}(1998)}]{normalmodes}
\bibinfo{author}{\bibfnamefont{A.~A.} \bibnamefont{Svidzinsky}}
  \bibnamefont{and} \bibinfo{author}{\bibfnamefont{A.~L.}
  \bibnamefont{Fetter}}, \bibinfo{journal}{Physical Review A}
  \textbf{\bibinfo{volume}{58}}, \bibinfo{pages}{3168} (\bibinfo{year}{1998}).

\bibitem[{\citenamefont{O'Dell and Eberlein}(2007)}]{vi1}
\bibinfo{author}{\bibfnamefont{D.~H.~J.} \bibnamefont{O'Dell}}
  \bibnamefont{and} \bibinfo{author}{\bibfnamefont{C.}~\bibnamefont{Eberlein}},
  \bibinfo{journal}{Physical Review A} \textbf{\bibinfo{volume}{75}},
  \bibinfo{pages}{013604} (\bibinfo{year}{2007}).

\bibitem[{\citenamefont{Dalfovo and Modugno}(2000)}]{michele}
\bibinfo{author}{\bibfnamefont{F.}~\bibnamefont{Dalfovo}} \bibnamefont{and}
  \bibinfo{author}{\bibfnamefont{M.}~\bibnamefont{Modugno}},
  \bibinfo{journal}{Physical Review A} \textbf{\bibinfo{volume}{61}},
  \bibinfo{pages}{023605} (\bibinfo{year}{2000}).

\bibitem[{\citenamefont{Teles et~al.}(2013)\citenamefont{Teles, dos Santos,
  Caracanhas, and Bagnato}}]{rpteles1}
\bibinfo{author}{\bibfnamefont{R.~P.} \bibnamefont{Teles}},
  \bibinfo{author}{\bibfnamefont{F.~E.~A.} \bibnamefont{dos Santos}},
  \bibinfo{author}{\bibfnamefont{M.~A.} \bibnamefont{Caracanhas}},
  \bibnamefont{and} \bibinfo{author}{\bibfnamefont{V.~S.}
  \bibnamefont{Bagnato}}, \bibinfo{journal}{Physical Review A}
  \textbf{\bibinfo{volume}{87}}, \bibinfo{pages}{033622}
  (\bibinfo{year}{2013}).

\bibitem[{\citenamefont{de~Lima~Henn}(2008)}]{emanuel}
\bibinfo{author}{\bibfnamefont{E.~A.} \bibnamefont{de~Lima~Henn}}, Ph.D.
  thesis, \bibinfo{school}{Instituto de F{\'\i}sica de S{\~a}o Carlos,
  Universidade de S{\~a}o Paulo}, \bibinfo{address}{S{\~a}o Carlos}
  (\bibinfo{year}{2008}).

\bibitem[{\citenamefont{Dennis et~al.}(2013)\citenamefont{Dennis, Hope, and
  Johnsson}}]{xmds}
\bibinfo{author}{\bibfnamefont{G.~R.} \bibnamefont{Dennis}},
  \bibinfo{author}{\bibfnamefont{J.~J.} \bibnamefont{Hope}}, \bibnamefont{and}
  \bibinfo{author}{\bibfnamefont{M.~T.} \bibnamefont{Johnsson}},
  \bibinfo{journal}{Computer Physics Communications}
  \textbf{\bibinfo{volume}{184}}, \bibinfo{pages}{201} (\bibinfo{year}{2013}).

\bibitem[{\citenamefont{Ketterle}(2001)}]{MIT}
\bibinfo{author}{\bibfnamefont{W.}~\bibnamefont{Ketterle}},
  \bibinfo{journal}{MIT Physics Annual} pp. \bibinfo{pages}{44--49}
  (\bibinfo{year}{2001}).

\bibitem[{\citenamefont{Henn et~al.}(2009{\natexlab{a}})\citenamefont{Henn,
  Seman, Ramos, Caracanhas, Castilho, Ol{\'\i}mpio, Roati, Magalh{\~a}es,
  Magalh{\~a}es, and Bagnato}}]{obv}
\bibinfo{author}{\bibfnamefont{E.~A.~L.} \bibnamefont{Henn}},
  \bibinfo{author}{\bibfnamefont{J.~A.} \bibnamefont{Seman}},
  \bibinfo{author}{\bibfnamefont{E.~R.~F.} \bibnamefont{Ramos}},
  \bibinfo{author}{\bibfnamefont{M.}~\bibnamefont{Caracanhas}},
  \bibinfo{author}{\bibfnamefont{P.}~\bibnamefont{Castilho}},
  \bibinfo{author}{\bibfnamefont{E.~P.} \bibnamefont{Ol{\'\i}mpio}},
  \bibinfo{author}{\bibfnamefont{G.}~\bibnamefont{Roati}},
  \bibinfo{author}{\bibfnamefont{D.~V.} \bibnamefont{Magalh{\~a}es}},
  \bibinfo{author}{\bibfnamefont{K.~M.~F.} \bibnamefont{Magalh{\~a}es}},
  \bibnamefont{and} \bibinfo{author}{\bibfnamefont{V.~S.}
  \bibnamefont{Bagnato}}, \bibinfo{journal}{Physical Review A}
  \textbf{\bibinfo{volume}{79}}, \bibinfo{pages}{043618}
  (\bibinfo{year}{2009}{\natexlab{a}}).

\bibitem[{\citenamefont{Henn et~al.}(2009{\natexlab{b}})\citenamefont{Henn,
  Seman, Roati, Magalh{\~a}es, and Bagnato}}]{turb}
\bibinfo{author}{\bibfnamefont{E.~A.~L.} \bibnamefont{Henn}},
  \bibinfo{author}{\bibfnamefont{J.~A.} \bibnamefont{Seman}},
  \bibinfo{author}{\bibfnamefont{G.}~\bibnamefont{Roati}},
  \bibinfo{author}{\bibfnamefont{K.~M.~F.} \bibnamefont{Magalh{\~a}es}},
  \bibnamefont{and} \bibinfo{author}{\bibfnamefont{V.~S.}
  \bibnamefont{Bagnato}}, \bibinfo{journal}{Physical Review Letters}
  \textbf{\bibinfo{volume}{103}}, \bibinfo{pages}{045301}
  (\bibinfo{year}{2009}{\natexlab{b}}).

\bibitem[{\citenamefont{Chevy et~al.}(2002)\citenamefont{Chevy, Madison, and
  Dalibard}}]{vor1}
\bibinfo{author}{\bibfnamefont{F.}~\bibnamefont{Chevy}},
  \bibinfo{author}{\bibfnamefont{K.~W.} \bibnamefont{Madison}},
  \bibnamefont{and} \bibinfo{author}{\bibfnamefont{J.}~\bibnamefont{Dalibard}},
  \bibinfo{journal}{Physical Review Letters} \textbf{\bibinfo{volume}{85}},
  \bibinfo{pages}{2223} (\bibinfo{year}{2002}).

\bibitem[{\citenamefont{Anderson and Haljan}(2000)}]{vp}
\bibinfo{author}{\bibfnamefont{B.~P.} \bibnamefont{Anderson}} \bibnamefont{and}
  \bibinfo{author}{\bibfnamefont{P.~C.} \bibnamefont{Haljan}},
  \bibinfo{journal}{Physical Review Letters} \textbf{\bibinfo{volume}{85}},
  \bibinfo{pages}{2857} (\bibinfo{year}{2000}).

\end{thebibliography}

\end{document}